\documentstyle[sprocl]{article}
\def\lref#1#2{{\bibitem{#1}#2}}

\def\np#1#2#3{Nucl. Phys. {\bf B#1} (#2), #3}
\def\pl#1#2#3{Phys. Lett. {\bf #1B} (#2), #3}
\def\prl#1#2#3{Phys. Rev. Lett. {\bf #1} (#2), #3}
\def\pr#1#2#3{Phys. Rev. {\bf #1} (#2), #3}
\def\prd#1#2#3{Phys. Rev. {\bf D#1} (#2), #3}
\def\aph#1#2#3{Ann. Phys. {\bf #1} (#2), #3}

\def\cmp#1#2#3{Comm. Math. Phys. {\bf #1} (#2), #3}

\def\jhep#1#2#3{JHEP {\bf#1}(#2), #3}

\def\atmp#1#2#3{Adv.~Theor.~Math.~Phys.{\bf #1} (#2), #3}
\def\ap#1#2#3{Ann.~Phys. {\bf #1} (#2), #3}
\def\IB{\relax\hbox{$\inbar\kern-.3em{\rm B}$}}
\def\IC{{\bf C}}
\def\ID{\relax\hbox{$\inbar\kern-.3em{\rm D}$}}
\def\IE{\relax\hbox{$\inbar\kern-.3em{\rm E}$}}
\def\IF{\relax\hbox{$\inbar\kern-.3em{\rm F}$}}
\def\IG{\relax\hbox{$\inbar\kern-.3em{\rm G}$}}
\def\IGa{\relax\hbox{${\rm I}\kern-.18em\Gamma$}}
\def\IH{\relax{\rm I\kern-.18em H}}
\def\IK{\relax{\rm I\kern-.18em K}}
\def\IL{\relax{\rm I\kern-.18em L}}
\def\IP{\relax{\rm I\kern-.18em P}}
\def\IR{{\bf R}}
\def\IZ{\relax\ifmmode\mathchoice{
\hbox{\cmss Z\kern-.4em Z}}{\hbox{\cmss Z\kern-.4em Z}}
                 {\lower.9pt\hbox{\cmsss Z\kern-.4em Z}}
                {\lower1.2pt\hbox{\cmsss Z\kern-.4em Z}}
                          \else{\cmss Z\kern-.4em Z}\fi}
\def\II{\relax{\rm I\kern-.18em I}}

\def\ndt{{\noindent}}

\def\CA{{\cal A}}

\def\CD{{\cal D}}
\def\CE{{\cal E}}
\def\CF{{\cal F}}
\def\CG{{\cal G}}
\def\CH{{\cal H}}
\def\CI{{\cal I}}

\def\CM{{\cal M}}
\def\CN{{\cal N}}
\def\CO{{\cal O}}

\def\CS{{\cal S}}
\def\CT{{\cal T}}
\def\CU{{\cal U}}

\def\CW{{\cal W}}

\def\CZ{{\cal Z}}

\def\p{{\partial}}
\def\pb{\bar{\p}}

\def\nb{\bar{n}}

\def\wb{\bar{w}}

\def\tb{\bar{t}}
\def\zb{\bar{z}}

\def\inv{^{\scriptscriptstyle{-1}}}
\def\half{{\scriptscriptstyle{1 \over 2}}}

\def\Tr{{\rm Tr}}
\def\Id{{\rm Id}}

\def\inbar{\,\vrule height1.5ex width.4pt depth0pt}
\font\cmss=cmss10 \font\cmsss=cmss10 at 7pt

\def\a{{\alpha}}
\def\ap{{{\a}^{\prime}}}
\def\b{{\beta}}
\def\d{{\delta}}
\def\g{{\gamma}}
\def\e{{\epsilon}}
\def\z{{\zeta}}
\def\ve{{\varepsilon}}
\def\vf{{\varphi}}
\def\m{{\mu}}
\def\n{{\nu}}

\def\k{{\kappa}}
\def\l{{\lambda}}
\def\s{{\sigma}}
\def\t{{\theta}}
\def\o{{\omega}}
\def\nct{{noncommutative \,}}
\def\exc{\ndt$\bullet$\small}
\def\be{\begin{equation}}
\def\ee{\end{equation}}
\def\bea{\begin{eqnarray}}
\def\eea{\end{eqnarray}}
\def\crt{\nonumber \\}
\def\eqn#1#2{{\be #2 \label{#1} \ee}}

\def\ctn#1{{{\cite{#1}}}}
\def\rfn#1{{({\ref{#1}})}}
\def\i{{\bf i}}
\begin{document}
\title{{\Large TRIESTE LECTURES}\\ ON SOLITONS IN \\ NONCOMMUTATIVE GAUGE THEORIES}
\author{Nikita A. Nekrasov\footnote{also at
Institute for Theoretical and Experimental Physics, 117259 Moscow,
Russia}}\address{IHES, Le Bois-Marie, 35 route de Chartres,
Bures-sur-Yvette, F-91440, France\\} \maketitle \abstracts{We
present a pedagogical introduction into noncommutative gauge
theories, their stringy origin, and non-perturbative effects,
including monopole and instantons solutions.}
\section{Introduction}

Recently there has been a revival of interest in noncommutative
gauge theories\ctn{snyder,connes,doplich}. They are interesting
examples of nonlocal field theories which in the certain limit (of
large noncommutativity) become essentially equivalent to the large
$N$ ordinary gauge theories \ctn{planar,filk}; certain
supersymmetric versions of \nct gauge theories  arise as ${\ap}
\to 0$ limit of theories on Dp-branes in the presence of
background $B$-field \ctn{douglashull,witsei}; the related
theories arise in Matrix compactifications with $C$-field turned
on \ctn{cds}; finally, noncommutativity is in some sense an
intrinsic feature of the open string field theory
\ctn{wtnc,k,volker}.

A lot of progress has been recently achieved in the analysis of
the classical solutions of the \nct gauge theory on the \nct
versions of Minkowski or Euclidean spaces. The first explicit
solutions and their moduli where analyzed in \ctn{neksch} where
instantons in the four dimensional \nct gauge theory (with
self-dual noncommutativity) were constructed. These instantons
play an important role in the construction of the discrete light
cone quantization of the M-theory fivebrane \ctn{abkss,abs}, and
they also gave a hope of giving an interpretation in the physical
gauge theory language of the torsion free sheaves which appear in
various interpretations of D-brane states
\ctn{avatars,harveymoore}, in particular those responsible for the
enthropy of black holes realized via D5-D1 systems
\ctn{vafastrominger}, and also enter the S-duality invariant
partition functions of ${\CN}=4$ super-Yang-Mills theory
\ctn{vafawitten}. One can also relax the self-duality assumption
on the noncommutativity. The construction of \ctn{neksch} easily
generalizes\ctn{nekinst} to the general case ${\t} \neq 0$ (see
also \ctn{fkiii,fkiv}). In addition to the instantons (which are
particles in 4+1 dimensional theory), which represent the D0-D4
system, the monopole-like solutions were found\ctn{grossnek} in
U(1) gauge theory in 3+1 dimensions. The latter turn out to have a
string attached to them. The string with the monopole at its end
are the \nct field theory realization of the D3-D1 system, where
D1 string ends on the D3 brane and bends at some specific angle
towards the brane. One can also find the solutions describing the
string itself \ctn{alexios}, \ctn{grossneki}, both the BPS and in
the non-BPS states; also the dimensionally reduced solutions in
2+1 dimensions \ctn{grossneki,amgs}, describing the D0-D2 systems;
finite length strings, corresponding to $U(2)$ monopoles
\ctn{grossnekii}.

The solitonic strings described in these lectures all carry
magnetic fluxes. Their S-dual electric flux strings represent
fundamental strings located nearby the D-branes with space-time
noncommutativity. Also we do not consider theories on the \nct
compact manifolds, like tori. Some classical and quantum aspects
of the Yang-Mills theory on the \nct tori are analyzed in
\ctn{nctori}.

These lectures are organized as follows. The section $2$ contains
a pedagogical introduction into \nct gauge theories. Section $3$
explains how \nct gauge theories arise as $\ap \to 0$ limits of
open string theory. The section $4$ constructs instantons in \nct
gauge theory on ${\bf R}^4$ for any group $U(N)$. The section $5$
presents explicit formulae for the $U(1)$ gauge group. The section
$6$ presents monopole solutions in $U(1)$ and $U(2)$ \nct gauge
theories. The section $7$ is devoted to some historic remarks. The
format of these lectures does not allow to cover all interesting
aspects of the \nct gauge theories, both classical and quantum,
and their relations to string/M-theory, and to large $N$ ordinary
gauge theories. We refer the interested readers to the
review\ctn{review} which will address all these issues in greater
detail.

{\exc Exercises are printed with the help of small fonts.} 
The symbol $\i$ denotes 
$\sqrt{-1}$, not to be confused with the space-time index $i$. The letter
${\t}$ will be used only for the Poisson tensor ${\t}^{ij}$, or its 
components, while
gauge theory theta angle will be denoted by ${\vartheta}$. The symbol $*$ denotes Hodge star, while $\star$ stands for star-products.

\section{Noncommutative Geometry and Noncommutative Field Theory}

\subsection{A brief mathematical introduction}

It has been widely appreciated by the mathematicians (starting
with the seminal works of  Gelfand, Grothendieck, and von Neumann)
that the geometrical properties of a space $X$ are encoded in the
properties  of the commutative algebra $C(X)$ of the continuous
functions $f: X \to {\IC}$ with the ordinary rules of point-wise
addition and multiplication: $(f + g)(x) = f(x) + g(x), f\cdot g
(x) = f(x) g(x)$.

More precisely, $C(X)$ knows only about the topology of $X$, but
one can refine the definitions and look at the algebra
$C^{\infty}(X)$ of the smooth functions or even at the DeRham
complex ${\Omega}^{\cdot}(X)$ to decipher the geometry of $X$.

The algebra ${\CA} = C(X)$ is clearly associative, commutative and
has a unit (${\bf 1} (x) = 1$). It also has an involution, which
maps a function to its complex conjugate: $f^{\dagger} (x) =
\overline{f(x)}$.

The points $x$ of $X$ can be viewed in two ways: as maximal ideals
of ${\CA}$: $f \in {\CI}_{x} \Leftrightarrow f(x) = 0$; or as the
irreducible (and therefore one-dimensional for ${\CA}$ is
commutative) representations of ${\CA}$: $R_x (f) = f(x)$, $R_x
\approx {\IC}$.

The vector bundles over $X$ give rise to projective modules over
${\CA}$. Given a bundle $E$ let us consider the space ${\CE} =
{\Gamma} (E)$ of its sections. If $f \in \CA$ and ${\s} \in {\CE}$
then clearly $f {\s} \in {\CE}$. This makes ${\CE}$ a
representation of $\CA$, i.e. a module. Not every module over
$\CA$ arises in this way. The vector bundles over topological
spaces have the following remarkable property, which is the
content of Serre-Swan theorem: for every vector bundle $E$ there
exists another bundle $E^{\prime}$ such that the direct sum $E
\oplus E^{\prime}$ is a trivial bundle $X \times {\IC}^{N}$ for
sufficiently large $N$. When translated to the language of modules
this property reads as follows: for the module ${\CE}$ over $A$
there exists another module ${\CE}^{\prime}$ such that ${\CE}
\oplus {\CE}^{\prime} = F_{N} = {\CA}^{\oplus N}$. We have denoted
by $F_{N} = {\CA} \otimes_{\IC} {\IC}^{N}$ the so-called free
module over $\CA$ of rank $N$. Unless otherwise stated the symbol
${\otimes}$ below will be used for tensor products over ${\IC}$.
The modules with this property are called {\it projective}. The
reason for them to be called in such a way is that ${\CE}$ is an
image of the free module $F_{N}$ under the projection which is
identity on ${\CE}$ and zero on ${\CE}^{\prime}$. In other words,
for each projective module ${\CE}$ there exists $N$ and an
operator $P \in {\rm Hom} (F_{N}, F_{N})$, such that $P^2 = P$,
and ${\CE} = P \cdot F_{N}$.

Noncommutative geometry relaxes the condition that $\CA$ must be
commutative, and develops a geometrical intuition about the
noncommutative associative algebras with anti-holomorphic
involution $^{\dagger}$ (${\IC}^*$-algebras).

In particular, the notion of vector bundle over $X$ is replaced by
the notion of the projective module over $\CA$. Now, when $\CA$ is
noncommutative, there are two kinds of modules: left and right
ones. The left $\CA$-module is the vector space $M_{l}$ with the
operation of left multiplication by the elements of the algebra
$\CA$: for $m \in M_l$ and  $a \in \CA$ there must be an element
$a m \in M_l$, such that for $a_1, a_2$: $a_1 (a_2 m) = (a_1 a_2)
m$. The definition of the right $\CA$-module $M_r$ is similar: for
$m \in M_r$ and  $a \in \CA$ there must be an element $m a\in
M_r$, such that for $a_1, a_2$: $(m a_1) a_2  = m (a_1 a_2)$. The
free module $F_N = \CA \oplus \ldots_{\scriptscriptstyle N \, \rm
times}\oplus \CA = \CA \otimes {\IC}^{N}$ is both left and right
one. The projective $\CA$-modules are defined just as in the
commutative case, except that for the left projective $\CA$-module
${\CE}$ the module ${\CE}^{\prime}$, such that ${\CE} \oplus
{\CE}^{\prime} = F_N$, also must be left, and similarly for the
right modules.

The manifolds can be mapped one to another by means of smooth
maps: $g : X_1 \to X_2$. The algebras of smooth functions  are
mapped in the opposite way: $g^* : C^{\infty}(X_2) \to
C^{\infty}(X_1)$, $g^* (f) (x_1) = f ( g(x_1))$. The induced map
of the algebras is the algebra homomorphism: $$g^* (f_1 f_2) =
g^*(f_1) g^*(f_2), \, g^* (f_1 + f_2) = g^* ( f_1) + g^* (f_2)$$

Naturally, the smooth maps between two manifolds are replaced by
the homomorphisms of the corresponding algebras. In particular,
the maps of the manifold to itself form the associative algebra
$Hom ({\CA}, {\CA})$. The diffeomorphisms would correspond to the
invertible homomorphisms, i.e. automorphisms $Aut ({\CA})$. Among
those there are internal, generated by the invertible elements of
the algebra: $$ a \mapsto g^{-1} a g $$ The infinitesimal
diffeomorphisms of the ordinary manifolds are generated by the
vector fields $V^i {\p}_i$, which
 differentiate
functions, $$f \mapsto f + {\ve} V^i {\p}_i f $$ In the
noncommutative setup the vector field is replaced by the
derivation of the algebra $V \in Der ({\CA})$: $$ a \mapsto a +
{\ve} V(a), \quad V( a) \in {\CA}$$ and the condition that $V(a)$
generates an infinitesimal homomorphism reads as: $$V (a b) = V
(a) b + a V(b)$$ which is just the definition of the derivation.
Among various derivations there are internal ones, generated by
the elements of the algebra itself: $$V_{c}(a) = [ a, c] : = a c -
c a, \quad c \in {\CA}$$ These infinitesimal diffeomorphisms are
absent in the commutative setup, but they have close relatives in
the case of Poisson manifold $X$.

\subsection{Flat \nct space}

The basic example of the noncommutative algebra which will be
studied here is the enveloping algebra of the Heisenberg algebra.
Consider the Euclidean space ${\bf R}^{d}$ with coordinates $x^i$,
$i=1, \ldots, d$. Suppose a constant antisymmetric matrix
${\t}^{ij}$ is fixed. It defines a Poisson bi-vector field
${\half}{\t}^{ij} {\p}_i \wedge {\p}_j$ and therefore the
noncommutative associative product on ${\bf R}^{d}$. The
coordinate functions $x^i$ on the deformed noncommutative manifold
will obey the following commutation relations: \eqn{cmrl}{[x^i,
x^j] = {\i} {\t}^{ij}\ , } We shall call the algebra ${\CA}_{\t}$
(over ${\IC}$) generated by the $x^i$ satisfying \rfn{cmrl},
together with  convergence conditions on the allowed expressions
of the $x^i$ -- the noncommutative space-time. The algebra
${\CA}_{\t}$ has an involution $a \mapsto a^{\dagger}$ which acts
as a complex conjugation on the central elements $\left( {\l}
\cdot {\bf 1} \right)^{\dagger} = {\bar\l} \cdot {\bf 1}, \, \l
\in {\IC}$ and preserves $x^i$: $(x^i)^{\dagger} = x^i$. The
elements of ${\CA}_{\t}$ can be identified with ordinary
complex-valued functions on ${\bf R}^d$, with the product of two
functions $f$ and $g$ given by the Moyal formula (or star
product): \eqn{myl}{f \star g \, (x)= {\exp} \left[ {{\i} \over 2}
{\t}^{ij} {{\p}\over{\p x_{1}^{i}}} {{\p}\over{\p x_{2}^{j}}}
\right] f (x_{1}) g (x_{2}) \vert_{x_{1} = x_{2} = x}\ .}

\subsubsection{Fock space formalism.}

By an orthogonal change of coordinates we can map the Poisson
tensor ${\t}_{ij}$ onto its canonical form: $$ x^i \mapsto z_a,
{\zb}_{a}, \quad a = 1, \ldots, r\ ; \quad y_{b}, \quad b = 1,
\ldots, d- 2r ,$$ so that: \eqn{ncm}{[y_a,y_b]= [y_b, z_a] = [y_b,
{\zb_a}] = 0, \quad [z_{a}, {\zb}_{b}] = -2{\t}_{a}{\d}_{ab} ,
{\quad} {\t}_{a}
> 0} $$ ds^2 =dx^2_i+ dy_b^2 = dz_a d{\zb}_{a} + dy_b^2 . $$
{}Since $z (\zb)$ satisfy (up to a constant) the commutation
relations of creation (annihilation) operators we can identify
functions $f(x,y)$ with  the functions of the $y_a$ valued in the
space of operators acting in the Fock space ${\CH}_r$ of $r$
creation and annihilation operators: \eqn{fock}{{\CH}_r =
\bigoplus_{\vec n}\, {\IC} \, \vert n_1, \ldots, n_r \rangle} $$
c_{a} = {1\over \sqrt{2{\t}_a}} {\zb}_a , \quad c^{\dagger}_a =
{1\over \sqrt{2\t_a}} z_{a}, \,  [c_{a}, c^{\dagger}_{b}] =
{\d}_{ab} $$ $$ c_a \vert {\vec n} \rangle = \sqrt{n_{a}} \vert
{\vec n}-1_{a} \rangle, \quad c_a^{\dagger} \vert {\vec n} \rangle
= \sqrt{n_{a} + 1} \vert {\vec n} + 1_a \rangle $$ Let ${\hat
n}_{a} = c^{\dagger}_{a} c_{a}$ be the $a$'th number operator.

The Hilbert space ${\CH}_r$ is the example of left projective
module over the algebra ${\CA}_{\t}$. Indeed, consider the element
$P_{0} = \vert \vec 0 \rangle \langle \vec 0 \vert \sim {\exp} -
\sum_{a} {{z_a {\zb}_a}\over{{\t}_a}}$. It obeys $P_0^2 = P_0$,
i.e. it is a projector. Consider the rank one free module $F_{1} =
{\CA}_{\t}$ and let us consider its left sub-module, spanned by
the elements of the form: $f \star P_0$. As a module it is clearly
isomorphic to ${\CH}_r$, isomorphism being: ${\vert \vec n
\rangle} \mapsto {\vert \vec n \rangle \langle
 \vec 0 \vert}$. It is projective module, the complementary
module being ${\CA}_{\t} ( 1 - P_0) \subset {\CA}_{\t}$.

{\ndt}The procedure that maps ordinary commutative functions onto
operators in the Fock space acted on by $z_a, {\zb}_a$ is called
Weyl ordering and is defined by: \eqn{wlor}{f(x) \mapsto {\hat
f(z_a, {\bar z}_a)} =  \int f(x) \, {{{\rm d}^{2r} x \,\, {\rm
d}^{2r} p }\over{(2{\pi})^{2r}}} \,
  \,  e^{ {\i} \left( {\bar p}_a z_a + p_{a} {\zb}_a -  p \cdot x \right)}.}
{\exc Show that if $f \mapsto \hat f, g \mapsto \hat g$ then
$f\star g \mapsto \hat f \hat g$.}

\subsubsection{Integration over the \nct space}
Weyl ordering also allows to express the integrals of the ordinary
functions over the commutative space in terms of the traces of the
operators, corresponding to them: \eqn{intgr}{\int {\rm d}^{2r} x
\, f(x) = ( 2\pi \t)^{r} {\Tr}_{{\CH}_r} {\hat f},} as follows
immediately from \rfn{wlor}. Sometimes the integral reduces to the
boundary term. What is the boundary term in the noncommutative,
Fock space setup?

To be specific, let us consider the case $r=1$. The general case
follows trivially. Consider the integral \eqn{divr}{\int {\rm d}^2
x \, \left(  {\p}_1 \, f_1 + {\p}_2 \, f_2 \right) = (2{\pi}i)
{\Tr}_{\CH} \left( [{\hat f}_1 , x^2] + [ x^1, {\hat f}_2] \right)
= \pi \sqrt{2\t} {\Tr} \left( [ c, {\bf f}] - [c^{\dagger}, {\bf f
}^{\dagger}] \right)} ${\bf f} = {\hat f}_1 + i {\hat f}_2$. In
computing the trace \eqn{trc}{ {\Tr}_{\CH} [ c, {\bf f} ] =
\sum_{n} \langle n \vert [ c, {\bf f} ] \vert n \rangle \ ,}we get
naively zero, for the trace of a commutator usually vanishes. But
we should be careful, since the matrices are infinite and the
trace is an infinite sum. If we regulate it by restricting the sum
to $n \leq N$, then the matrix element $\langle N  \vert c \vert
N+1 \rangle \langle N+1 \vert {\bf f} \vert N \rangle$ is not
cancelled, so that the regularized trace is
\eqn{trcrg}{{\Tr}_{{\CH}_{N}} [ c, {\bf f} ]  = \sqrt{N+1} \langle
N+1 \vert {\bf f} \vert N \rangle} and similarly for
$c^{\dagger}$. Thus: \eqn{bndryt}{\oint_{\infty} f_1 {\rm d}x^2 -
f_2 {\rm d} x^1 = 2\pi \sqrt{2\t (N+1)} {\rm Re} \langle N+1 \vert
\, {\hat f}_1 + i {\hat f}_2 \, \vert N \rangle_{N \to \infty}}
{\exc Consider ${\bf f} = {1\over{c^{\dagger}c}} c^{\dagger}$.
Compute the integral \rfn{divr} directly and via \rfn{bndryt}.}

\ndt Let us conclude this section with the remark that \rfn{intgr}
defines the trace on a part of the algebra ${\CA}_{\t}$ consisting
of the trace class operators. They correspond to the space $L^{1}
({\bf R}^{2r}) $ of integrable functions. Even though the trace is
taken in the specific representation of ${\CA}_{\t}$ it is defined
intrinsically, thanks to \rfn{intgr}. In what follows we omit the
subscript ${\CH}_r$ in the notation for the trace on ${\CA}_{\t}$.

\subsubsection{Symmetries of the flat noncommutative space}

The algebra \rfn{cmrl} has an obvious symmetry: $x^i \mapsto x^i +
{\ve}^i$, with ${\ve}^i \in {\bf R}$. For invertible Poisson
structure ${\t}$, such that ${\t}_{ij} {\t}^{jk} = {\d}_{i}^{k}$
this symmetry is an example of the internal automorphism of the
algebra: \eqn{auto}{a \mapsto e^{ {\i}{\t}_{ij} {\ve}^{i} x^{j}} a
e^{-{\i}{\t}_{ij}{\ve}^{i}x^j}} In addition, there are rotational
symmetries which we shall not need.

\subsection{Gauge theory on  noncommutative space}

{}In an  ordinary gauge theory with gauge group $G$ the gauge
fields are connections in some principal $G$-bundle. The matter
fields are the sections of the vector bundles with the structure
group $G$. Sections of the noncommutative vector bundles are
elements of the projective modules over the algebra ${\CA}_{\t}$.

\subsubsection{Gauge fields, matter fields ....}
{}In the ordinary gauge theory the gauge field arises through the
operation of covariant differentiation of the sections of a vector
bundle.  In the \nct setup the situation is similar. Suppose $M$
is a projective module over ${\CA}_{\t}$. The connection
${\nabla}$ is the operator $$ {\nabla} : {\bf R}^d \times M \to M,
\quad {\nabla}_{\ve} (m) \in M, \qquad {\ve} \in {\bf R}^d, \, m
\in M \ ,  $$ where ${\bf R}^d$ denotes the commutative vector
space, the Lie algebra of the automorphism group generated by
\rfn{auto}. The connection is required to obey the Leibnitz rule:
\eqn{leibn}{{\nabla}_{\ve} ( a m_{\bf l} ) = {\ve}^i ( {\p}_i a )
m_{\bf l} + a {\nabla}_{\ve} m_{\bf l}} \eqn{reibn}{{\nabla}_{\ve}
( m_{\bf r} a ) = m_{\bf r} {\ve}^i ( {\p}_i a ) + (
{\nabla}_{\ve} m_{\bf r} ) a \ . } Here, \rfn{leibn} is the
condition for left modules, and \rfn{reibn} is the condition for
the right modules. {}As usual, one defines the curvature $F_{ij} =
[{\nabla}_i, {\nabla}_j]$ - the operator ${\Lambda}^2 {\bf R}^d
\times M \to M$ which commutes with the multiplication by $a \in
{\CA}_{\t}$. In other words, $F_{ij} \in {\rm End}_{\CA}(M)$. In
ordinary gauge theories the gauge fields come with gauge
transformations. In the \nct case the gauge transformations, just
like the gauge fields, depend on the module they act in. For the
module $M$ the group of gauge transformations ${\CG}_{M}$ consists
of the invertible endomorphisms of $M$ which commute with the
action of ${\CA}$ on $M$: $${\CG}_{M} = {\rm GL}_{{\CA}_{\t}}
(M)$$Its Lie algebra ${\rm End}_{\CA} (M)$ will also be important
for us.

All the discussion above can be specified to the case where the
module has a Hermitian inner product, with values in ${\CA}_{\t}$.

In addition to gauge fields gauge theories often have matter
fields.  One should distinguish two types of matter fields. First
of all, the elements ${\vf}$ of the module $M$ where $\nabla$
acts, can be used as the matter fields. Then ${\nabla}_i {\vf}$ is
the usual covariant derivative of the element of the module. This
is the \nct analogue of the matter fields in the fundamental
representation of the gauge group. Another possibility is to look
at the Lie algebra of the gauge group ${\rm End}_{{\CA}_{\t}}
(M)$. Its elements $\Phi$ (we shall loosely call them adjoint
Higgs fields) commute with the action of the algebra ${\CA}$ in
the module, but act nontrivially on the elements of the module
$M$: ${\vf } \mapsto {\Phi} \cdot {\vf} \in M$. In particular, one
can consider the commutators between the covariant derivatives and
the Higgs fields: $[{\nabla}_i, {\Phi}] \in {\rm
End}_{{\CA}_{\t}}(M)$. These commutators can be called the
covariant derivatives of the adjoint Higgs fields.

The important source of such matter fields is the dimensional
reduction of the higher dimensional theory. Then the components of
the covariant derivative operator in the collapse directions
become the adjoint Higgs fields in the reduced theory.

\subsubsection{Fock module and connections there.}

Recall that the algebra ${\CA}_{\t}$ for $d = 2r$ and
non-degenerate ${\t}$ has an important irreducible representation,
the left module ${\CH}_{r}$. Let us now ask, what kind of
connections does the module ${\CH}_{r}$ have?

By definition\rfn{leibn}, we are looking for a collection of
operators ${\nabla}_{i} : {\CH}_r \to {\CH}_r$, $i=1, \ldots, 2r$,
such that: $$ [ {\nabla}_i, a ] = {\p}_i a $$ for any $a \in
{\CA}$. Using the fact that ${\p}_i a = i{\t}_{ij} [ x^j, a] $ and
the irreducibility of ${\CH}_r$ we conclude that:
\eqn{cnfc}{{\nabla}_i = {\i} {\t}_{ij} x^j + {\k}_i, \qquad {\k}_i
\in {\IC}} If we insist on unitarity of ${\nabla}$, then $i {\k}_i
\in {\bf R}$. Thus, the space of all gauge fields suitable for
acting in the Fock module is rather thin, and is isomorphic to the
vector space ${\bf R}^{d}$ (which is canonically dual to the Lie
algebra of the automorphisms of ${\CA}_{\t}$). The gauge group for
the Fock module, again due to its irreducibility is simply the
group $U(1)$, which multiplies all the vectors in ${\CG}_r$ by a
single phase. In particular, it preserves ${\k}_i$'s, so they are
gauge invariant. {}It remains to find out what is the curvature of
the gauge field given by \rfn{cnfc}. The straightforward
computation of the commutators gives: \eqn{fccrvt}{F_{ij} = {\i}
{\t}_{ij}} i.e. all connections in the Fock module have the
constant curvature.

\subsubsection{Free modules and connections there.}

If the right (left) module $M$ is free, i.e. it is a sum of
several copies of the algebra ${\CA}_{\t}$ itself, then the
connection ${\nabla}_i$ can be written as $$ {\nabla}_i = {\p}_i +
A_i $$ where $A_i$ is the operator of the left (right)
multiplication by the matrix with ${\CA}_{\t}$-valued entries:
\eqn{conct}{{\nabla}_i m_{\bf l} = {\p}_i m_{\bf l} + m_{\bf l}
A_i, \, {\nabla}_i m_{\bf r} = {\p}_i m_{\bf r} + A_i m_{\bf r}}In
the same operator sense the curvature obeys the standard identity:
$$ F_{ij} = {\p}_i A_j - {\p}_j A_i + A_i A_j - A_j A_i \ . $$
{}Given a module $M$ over some algebra $\CA$ one can multiply it
by a free module ${\CA}^{\oplus N}$ to make it a module over an
algebra ${\rm Mat}_{N \times N}({\CA})$ of matrices with elements
from ${\CA}$. In the non-abelian gauge theory over ${\CA}$ we are
interested in projective  modules over ${\rm Mat}_{N \times
N}({\CA})$. If the algebra $\CA$ (or perhaps its subalgebra) has a
trace, ${\Tr}$, then the algebra ${\rm Mat}_{N \times N} ({\CA})$
has a trace given by the composition of a usual matrix trace and
${\Tr}$.

{}It is a peculiar property of the noncommutative algebras that
the algebras $\CA$ and ${\rm Mat}_{N \times N}({\CA})$ have much
in common. These algebras are called Morita equivalent and under
some additional conditions the gauge theories over $\CA$ and over
${\rm Mat}_{N\times N}({\CA})$ are also equivalent. This
phenomenon is responsible for the similarity between the "abelian
noncommutative" and "non-abelian commutative" theories.

\subsubsection{Observables from gauge fields}
Just like in the commutative case the difference of two
connections in the module $M$ is the operator from ${\rm
End}_{{\CA}_{\t}}(M)$. Thus we can write: \eqn{gnrcc}{{\nabla}_i =
{\i} {\t}_{ij} x^j + D_i, \quad D_i \in {\rm End}_{{\CA}_{\t}}(M)}
The curvature of the connection ${\nabla}_i$ is, therefore:
\eqn{gnrcf}{F_{ij} = [ {\nabla}_i , {\nabla}_j ] = {\i} {\t}_{ij} +
[D_i, D_j] \in {\rm End}_{{\CA}_{\t}}(M)} For free modules the
operators $D_i$ appear from the formulae \rfn{conct} if we
represent ${\p}_i$ as ${\i} {\t}_{ij} [ x^j, \cdot]$:
\eqn{back}{{\nabla}_i m_{\bf l} = {\i} {\t}_{ij} x^j m_{\bf r} +
m_{\bf r} D_i, \quad {\nabla}_i m_{\bf r} =  - m_{\bf r} {\i}
{\t}_{ij}x^j + D_{i}^{\dagger} m_{\bf r}}The relation of the
operators $D_i$ and the conventional gauge fields $A_i$ is
\eqn{backi}{D_i = - {\i} {\t}_{ij} x^j + A_i} Going back to the
generic case, we shall now describe a (overcomplete) set of gauge
invariant observables in the gauge theory on the module $M$. The
gauge transformations act on the operators $D_i$ ``locally'', i.e.
for $g \in {\rm GL}_{{\CA}_{\t}}(M)$: \eqn{ggtrnf}{D_i \mapsto
g^{-1} D_i g} Hence the spectrum of the operators $D_i$, or of any
analytic function of them is gauge invariant. In particular, the
following observables are the \nct analogues of the Wilson loops
in the ordinary gauge theory. Choose a contour on the ordinary
space ${\bf R}^{2r}$: ${\g}^i (t)$, $0 \leq t \leq 1$. Define an
operator \eqn{dcnt}{{\CU}_{\g} = P {\exp} \int_{0}^{1} \, {\rm d}t
\, D_i \, {\dot \g}^i (t)} which also transforms as in
\rfn{ggtrnf}. The following observable \eqn{wlsn}{{\CW}_{\g} =
{\Tr}_{M}\, {\CU}_{\g}} is gauge-invariant. It is typically a
distribution on the space of contours ${\g}$.  For example, for
the Fock module $$ {\CW}_{\g} = e^{{\k}_i \left( {\g}^i (1) -
{\g}^i (0) \right)} \times {\Tr}_{{\CH}_r} \ 1 $$ while for the
free module in the vacuum ($D_i = - {\i}{\t}_{ij}x^j$): $$ {\CW}_{\g}
= {\d}^{2r} \left( {\t}_{ij} \left( {\g}^j (1) - {\g}^j (0)
\right)\right)\, {\exp}\,  {\i}  \left[ \oint_{\g} {\t}_{ij} {\g}^i d
{\g}^j  \right]$$ The operators \rfn{wlsn} are closely related to
the ``\nct momentum carrying Wilson loops'' considered in
\ctn{wilsl}.

\subsection{Lagrangian,  and couplings}

{}In order to write down the Lagrangian for the gauge theory in
the commutative setup one needs to specify a few details about the
space-time and the gauge theory: the space-time metric
$G_{{\m}{\n}}$, gauge coupling $g_{\rm YM}$, theta angle
${\vartheta}$ and so on. The same is true for the \nct theory,
except that the parameters above are more restricted, for given
${\CA}$. In these lectures we shall be dealing with static field
configurations, in the theories in p+1 dimensions. We shall only
look at the potential energy for such configurations. It is given
by: \eqn{glagr}{{\CE}(A) = -{1\over{4g_{\rm YM}^2}} \sum \sqrt{G}
G^{ii^{\prime}} G^{jj^{\prime}} {\Tr} F_{ij} F_{i^{\prime}
j^{\prime}} \ .} If additional adjoint Higgs matter fields $\Phi$
are present then the \rfn{glagr} becomes: \eqn{lagint}{{\CE}(A,
{\Phi}) = {\CE} (A) + \sum  \sqrt{G} G^{ii^{\prime}} {\Tr}
{\nabla}_i {\Phi} {\nabla}_{i^{\prime}} {\Phi} + \ldots} These
formulae make sense for  the constant Euclidean metric $G_{ij}$
only. String theory allows more general backgrounds, where the
closed string metric $g$ and the $B$ field both are allowed to be
non-constant. They are presumably described by more abstract
techniques of \nct geometry\ctn{connes} which we shall not use in
these lectures.

{}We now proceed with the exposition of how the associative
algebras, their deformations, and gauge theories over them  arise
in the string theory.

\section{Noncommutative geometry and strings in background $B$-fields}

\subsection{Conventional strings and D-branes} Let us look at the
theory of open strings in the following closed string background:
Flat space $X$, metric $g_{ij}$, constant Neveu-Schwarz $B$-field
$B_{ij}$; Dp-branes are present, so that $B_{ij} \neq 0$ for some
$i,j$ along the branes. The presence of the Dp-branes means that
the $B_{ij}$ cannot be gauged away - if we try to get rid of
$B_{ij}$ by means of the gauge transformation $$ B \to B + d
{\Lambda}$${}then  we create a gauge field $A_i$ on the brane,
whose field strength $F_{ij}$ is exactly equal to $B_{ij}$.

\ndt The bosonic part of the worldsheet action of our string is
given by: \eqn{wsa}{S = {1\over{4{\pi}{\ap}}} \int_{\Sigma} \left(
g_{ij} {\p}_{a} x^i {\p}^a x^j - 2{\pi}{\i}{\ap} B_{ij} {\e}^{ab}
{\p}_a x^i {\p}_b x^j \right)} $$ {1\over{4{\pi}{\ap}}}
\int_{\Sigma}  g_{ij} {\p}_{a} x^i {\p}^a x^j - {{\i}\over 2}
\int_{{\p}{\Sigma}} B_{ij} x^{i} {\p}_t x^j $$ Here we denote by
${\p}_t$ the tangential derivative along the boundary
${\p}{\Sigma}$ of the worldsheet ${\Sigma}$ (which we shall
momentarily take to be the upper half-plane). We shall also need
the normal derivative ${\p}_{n}$.

\ndt The boundary conditions which follow from varying the action
\rfn{wsa} are\footnote{It was observed by S.~Shatashvili in 1995
right after the appearance of\ctn{polch} that the $B$-field (or
equivalently the constant electromagnetic field) as well as the
tachyon interpolates between the Dirichlet and von Neumann
boundary conditions, and as a consequence between different
solutions of the target space lagrangian of background independent
open string field theory\ctn{witbsft,sambsft}, and that by taking
some of the components of $B_{ij} \to \infty$ one creates lower
dimensional D-branes. Similar properties of the tachyon
backgrounds are now extensively
investigated\ctn{samger,kmm,samgerii,tachyons,tachyonsen,radu}}:
\eqn{bndc}{g_{ij} {\p}_{n} x^i + 2{\pi} i {\ap} B_{ij} {\p}_t x^j
\vert_{{\p}{\Sigma}} = 0}

\ndt Now, on the upper half-plane with the coordinate $z= t + iy$,
$y > 0$ we can compute the propagator: \eqn{prp}{\langle x^i (z)
x^j (w) \rangle = } $$- {\ap} \left[ g^{ij} {\rm log} \left( {{z -
w}\over{z - {\wb}}} \right) + G^{ij} {\rm log} \vert z - w \vert^2
+ {1\over{2{\pi}{\ap}}} {\t}^{ij} {\rm log} \left( {{z -
{\wb}}\over{{\zb} - w}} \right)  + D^{ij} \right]$$ where $D^{ij}$
is independent of $z,w$, \eqn{opmb}{G^{ij} = \left( {1\over{g +
2{\pi} {\a}^{\prime} B }}\right)^{ij}_{S}} $${\t}^{ij} = 2{\pi}
{\a}^{\prime} \left( {1\over{g + 2{\pi}{\a}^{\prime} B}}
\right)^{ij}_{A}$$ where $S$ and $A$ denote the symmetric and
antisymmetric parts respectively. The open string vertex operators
are given by the expressions like \eqn{opsvrt}{: f\left(x(t), \,
{\p}_t x(t) , \,  {\p}_{t}^2 x(t), \, {\ldots} \right): } where
everything is evaluated at $y = 0$.

\ndt  The  properties of the open strings are encoded in the
operator product expansion of the open string vertex operators. By
specifying \rfn{prp} at $y =0$ we get the expression for the
propagator of the boundary values of the coordinates $x^j(t)$:
\eqn{bnprp}{\langle x^i (t) x^j (s) \rangle = - {\a}^{\prime}
G^{ij} {\rm log} ( t - s)^2 +{{\i}\over 2} {\t}^{ij} {\e}( t - s)}
where ${\e}(t) = -1, 0, +1$ for $t < 0, t = 0 , t > 0$
respectively. From this expression we deduce : \eqn{cmtr}{[ x^i ,
x^j] : = T(x^i ( t- 0) x^j(t) - x^i ( t+0) x^j(t)) = i {\t}^{ij}}
It means that the end-points of the open strings live on the
noncommutative space where: $$ [ x^i, x^j ] = {\i} {\t}^{ij} $$ with
${\t}^{ij}$ being a constant antisymmetric matrix. Similarly, the
OPE of the vertex operators: $$ V_{p} (t) = : e^{{\i} p {\cdot} x}:
(t) $$ is given by : \eqn{ope}{V_{p} (t) V_{q} (s) = ( t - s)^{2
{\a}^{\prime} G^{ij} p_{i} q_{j}} e^{-{{\i}\over{2}}{\t}^{ij} p_{i }
q_{j}} V_{p + q} (s) }

{\ndt}Seiberg and Witten suggested\ctn{witsei} to consider the
limit ${\ap} \to 0$ with $G, {\t}$ being kept fixed. In this limit
the OPE \rfn{ope} goes over to the formula for the modified
multiplication law on the ordinary functions on a space with the
coordinates $x$: \eqn{swope}{V_{p} V_{q} =
e^{-{{\i}\over{2}}{\t}^{ij} p_{i} q_{j}} V_{p + q}} (the appearance
of the noncommutative algebra \rfn{swope} in the case of
compactification on a shrinking torus was observed earlier by
M.~Douglas and C.~Hull in \ctn{douglashull} ). The algebra defined
in \rfn{swope} is isomorphic to Moyal algebra \rfn{myl}. The
product \rfn{swope} is associative but clearly noncommutative.

Witten\ctn{wittach} remarked that the ${\ap} \to 0$ limit with
fixed $g_{ij}, {\t}^{ij}$ makes the algebra of open string vertex operators
(which is associative as the vertex algebra) to factorize into the
product of the associative algebra of the string zero modes and
the algebra of string oscillators (excited modes). This allows to
see some stringy effects already at the level of the \nct field
theories.
\subsection{Effective action}

Vertex operators \rfn{opsvrt} give rise to the space-time fields
${\Phi}_k$ propagating along the worldvolume of the Dp-brane.
Their effective Lagrangian is obtained by evaluating the disc
amplitudes with \rfn{opsvrt} inserted at the boundary of the disc:
\eqn{effac}{\int {\rm d}^{p+1}x \, \sqrt{{\det}G} \, {\Tr}
{\p}^{n_1} {\Phi}_1 {\p}^{n_2} {\Phi}_2 \ldots {\p}^{n_k} {\Phi}_k
\sim \qquad\qquad\qquad}$$\qquad\qquad\qquad \langle
\prod_{m=1}^{k} :{\bf P}_{m} \left( {\p}x (t_m) , {\p}^2 x (t_m) ,
\ldots \right) e^{ip_m \cdot x(t_m)}: \rangle$$ If we compare the
effective actions of the theory at ${\t} = 0$ and ${\t} \neq 0$
the difference is very simple to evaluate: one should only take
into account the extra phase factors from \rfn{ope}:
\eqn{eac}{\langle \prod_{m=1}^{k} :{\bf P}_{m} \left( {\p}x (t_m)
, {\p}^2 x (t_m) , \ldots \right) e^{{\i}p_m \cdot x(t_m)}:
\rangle_{G, {\t}} =\qquad\qquad\qquad } $$ e^{ - {{\i}\over{2}}
\sum_{n
> m} p_i^n {\t}^{ij} p_j^m  {\e} (t_{n} - t_{m})} \langle
\prod_{m=1}^{k} :{\bf P}_{m} \left( {\p}x (t_m) , {\p}^2 x (t_m) ,
\ldots \right) e^{{\i}p_m \cdot x(t_m)}: \rangle_{G, 0} $$ (the
exponent in \rfn{eac} is depends only on the cyclic order of the
operators, due to the antisymmetry of ${\t}$ and the momentum
conservation: $$\sum_{m} p^m = 0$$ Using \rfn{swope},\rfn{myl} we
can easily conclude that the terms like \rfn{effac} go over to the
terms like: \eqn{nceac}{\int {\rm d}^{p+1}x \, \sqrt{{\det}G} \,
{\Tr} \, {\p}^{n_1} {\Phi}_1 \star {\p}^{n_2} {\Phi}_2 \star
\ldots \star {\p}^{n_k} {\Phi}_k } This result holds even without
taking the Seiberg-Witten limit.

In general, the relation between the off-shell string field theory
action and the boundary operators correlation functions in the
wolrdsheet conformal theory depends on the type of string theory
we are talking about. In the case of bosonic string in the
framework of Witten-Shatashvili background independent open string
field theory this relation reads as\ctn{witbsft,sambsft}:
\eqn{bsft}{{\CS} = {\CZ} - {\b}^i {{\p}\over{{\p t^i}}} {\CZ}}
where ${\CZ} (t)$ is the generating function of the boundary
correlators: $$ {\CZ} (t) = \langle {\exp}\oint_{\p \Sigma} t^i
{\CO}_i \rangle $$ with ${\CO}_i$ running through some basis in
the space of the open string vertex operators, $t^i$ being the
corresponding couplings, and ${\b}^i$ the ${\b}$-function of the
coupling $t^i$: $$ {\b}^i = {\Lambda} {{d}\over{d {\Lambda}}}
t^i$$The relation\rfn{bsft} derived in \ctn{sambsft} is very
restrictive, for ${\b}^i$ vanish exactly where $d{\CS}$ does, and
allows to calculate the exact expression for ${\CS}$ up to two
space-time derivatives\ctn{samger,kmm}. For the superstring it
seems\ctn{ssbft} that ${\CS} = {\CZ}$. This approach was
successfully applied to the study of D-branes with the B-field
turned on \ctn{lorenzo,okuyama,tachyons}.

\subsection{Gauge theory from string theory}

\ndt{}Open strings carry gauge fields. This follows from the
presence in the spectrum of the allowed vertex operators of the
operator of the following simple form: \eqn{gfvo}{e_{i}(p) :{\p}_t
x^i e^{{\i} p \cdot x} : \, \longleftrightarrow \, A = A_{i} (x)
dx^i}{}It has (classical) dimension $1$ and deforms the worldsheet conformal
field theory as follows: \eqn{defac}{ S = S_{0} -
{\i} \oint_{{\p}{\Sigma}} A_{i}(x) {\p}_t x^i \, {\rm d}t}{}This deformation
has the following naive symmetry: \eqn{nas}{ {\d} A_{j} = {\p}_j
{\ve}} ${\ve}(x)$ is the tachyonic vertex operator. Let us see,
whether \rfn{nas} is indeed a symmetry. To this end we estimate
the ${\d}$-variation of the disc correlation function:
\eqn{dsc}{{\d} \int Dx \, \, e^{-S} = \langle i\oint {\p}_t {\ve}
dt \rangle_{S} = \langle i\oint {\p}_t {\ve} dt \rangle_{S_{0}}}
$$ - \langle \oint\oint dt ds A_j (x(t)) {\p}_t x^j (t) {\p}_{s}
{\ve}(s) \rangle_{S_{0}} + \ldots $$ Let us regularize \rfn{dsc}
by point splitting, i.e. by understanding the $s,t$ integration
with $s \neq t$ (of course, there are other regularizations, their
equivalence leads to important predictions concerning the
noncommutative gauge theory, see \ctn{witsei}). Then the total
$s$-derivative needs not to decouple. Instead, it gives: $$-
\langle \oint dt \int_{s \neq t} ds A_j (x(t)) {\p}_t x^j (t)
{\p}_{s} {\ve}(x(s)) \rangle_{S_{0}} = $$ $$ \langle \oint dt A_j
(x(t)) {\p}_t x^j (t) \left( {\ve} (x(t + 0)) - {\ve}(x(t-0))
\right) \rangle_{S_{0}} = $$ $$ = \langle \oint {\ve} \star A - A
\star {\ve} \rangle_{S_{0}} $$ This calculation shows that the
naive transformation \rfn{nas} must be supplemented by the
correction term: \eqn{actgt}{{\d} A_j = {\p}_j {\ve} + A_j \star
{\ve} - {\ve} \star A_j } The formula \rfn{actgt} is exactly the
gauge transformation of the gauge field in the gauge theory on the
noncommutative space. Similarly, the effective action for the
gauge fields in the presence of ${\t} \neq 0$ becomes that of the
Yang-Mills theory on the noncommutative space plus the corrections
which vanish in the ${\ap} \to 0$ limit. In what follows we shall
need the relation between various string theory moduli in the case
of D3-branes in the background of the constant $B$-field:

\subsubsection{Open and closed string moduli.}

We want to consider the decoupling limit of a D3-brane in the Type
IIB string theory in a background with a constant Neveu-Schwarz
B-field. Let us recall
 the relation of the parameters of the actions \rfn{glagr},
 \rfn{lagint}
and the string theory parameters, before  taking the
Seiberg-Witten limit \ctn{witsei}.

We start with the D3-brane whose worldvolume is occupying the 0123
directions, and turn on a $B$-field: \eqn{bfi}{ {1\over 2} B dx^1
\wedge dx^2} The indices $i,j$ below will run from $1$ to $3$. We
assume that the closed string metric $g_{ij}$ is flat, and the
closed string coupling $g_{s}$ is small. According to \ctn{witsei}
the gauge theory on the D3-brane is described by a Lagrangian,
which, when  restricted to time-independent fields, equals
\rfn{lagint} with the parameters $$ G_{ij}, {\t}^{ij}, g_{\rm
YM}^2 \ , $$ which are related to $$ g_{ij}, B_{ij}, g_{s} $$ via
\rfn{opmb} as follows: \eqn{wsrl}{G_{ij} = g_{ij} -
(2{\pi}{\a}^{\prime})^2 \left( B g^{-1} B \right)_{ij},  \,
{\t}^{ij} = -(2{\pi}{\a}^{\prime})^2 \left( {1\over{g +
2{\pi}{\a}^{\prime} B}} B {1\over{g - 2{\pi}{\a}^{\prime}B}}
\right)^{ij}} $$ g_{\rm YM}^2 = 4{\pi}^2 g_{s} \left( {\rm det}
\left( 1 + 2{\pi}{\a}^{\prime} g^{-1}B \right)\right)^{\half}\ .
$$ Suppose the open string metric is Euclidean: $G_{ij} =
{\d}_{ij}$, then \rfn{bfi},\rfn{wsrl} imply: \eqn{clstr}{g =
dx_3^2 + {{(2{\pi}{\a}^{\prime})^2}\over{(2{\pi}{\a}^{\prime})^2 +
{\t}^2}} \left( dx_1^2 + dx_2^2 \right), \qquad B =
{{\t}\over{(2{\pi}{\a}^{\prime})^2  + {\t}^2}}\ , } and
\eqn{clscpl}{g_{s} = {{g_{\rm YM}^2}\over{2{\pi}}}
{{{\a}^{\prime}}\over{\sqrt{(2{\pi} {\a}^{\prime})^2 + {\t}^2}}}\
. } The Seiberg-Witten limit is achieved by taking ${\ap} \to 0$
with $G, {\t}, g_{\rm YM}^2$ kept fixed. In this limit the
effective action of the D3-brane theory will become that of the
(super)Yang-Mills theory on a noncommutative space ${\CA}_{\t}$.
The relevant part of the energy functional is: \eqn{nwenrg}{{\CE}
= {1\over{2g_{\rm YM}^2}} \int  \,
 {\Tr} \left( B_i \star B_i + {\nabla}_{i} {\Phi} \star
{\nabla}_i {\Phi} \right) \  ,} where
 \eqn{mgnf}{B_i = {i\over 2}
{\ve}_{ijk} F_{jk}\ .} The fluctuations of the D3-brane in some
distinguished transverse direction are described by the dynamics
of the Higgs field. Of course the true D3-brane theory will have
six adjoint Higgs fields, one per transverse direction. We are
looking at one of them. Also notice that the expression for the
energy \rfn{nwenrg} goes over to the case of $N$ D3-branes, one
simply replaces ${\CA}_{\t}$ by ${\rm Mat}_{N} ({\CA}_{\t})$

\subsection{Noncommutative geometry from topological string theory}

String theory can be well-defined even in the presence of
non-constant $B$-field. The absolute minima of the $B$-field
energy are achieved on the flat $B$-fields, i.e. if $dB = 0$. In
this case, if in addition the tensor $B_{ij}$ is invertible one
can define the inverse tensor ${\t}^{ij} = (B^{-1})^{ij}$ which
will obey Jacobi identity ${\t} {\p} {\t} =0$. It is not widely
appreciated in the string theory literature that a string can
actually propagate in the background of the non necessarily
invertible Poisson bi-vector ${\t}^{ij}$ (we leave the question of
whether the Poisson property can also be relaxed to future
investigations).

We shall now describe a version of a string theory, which is
perfectly sensible even for the non-constant Poisson bi-vector
field ${\t}^{ij}$, not necessarily invertible. To do this we shall
start with the bosonic string action with the target space $X$ and
then take the ${\ap} \to 0$ limit {\it a l\`a} Seiberg and Witten.
For simplicity we make an analytic continuation ${\t}^{ij} \to {\i} {\t}^{ij}$(cf.\ctn{k,cf}).

{\ndt}Take the action \rfn{wsa} not assuming that $g_{ij}, B_{ij}$
are constant and rewrite it in the first order form: \eqn{wsac}{S
= {1\over{4\pi\ap}} \int_{\Sigma} \left( g_{ij} {\p}_{a} x^i
{\p}^a x^j - 2{\pi}{\i}{\a}^{\prime} B_{ij} {\e}^{ab} {\p}_a x^i
{\p}_b x^j \right) \leftrightarrow} $$ \int p_{i} \wedge {\rm d}
x^i - {\pi}{\ap} \, G^{ij} p_{i} \wedge * p_{j} -
{1\over 2}{\t}^{ij} p_{i}\wedge p_{j} $$ where $$ 2{\pi}{\ap} G +
{\t} = 2{\pi}{\ap} \left( g + 2{\pi}{\ap} B \right)^{-1} $$ Now
suppose we take ${\ap} \to 0$ limit keeping ${\t}$ and $G$ fixed.
The remaining part $\int p \wedge {\rm d} x + {\half} {\t} p
\wedge p$ of the action \rfn{wsac} immediately exhibits an
enhancement of the (gauge) symmetry\ctn{cf}: \eqn{gsm}{p_i \mapsto
p_i - {\rm d} {\l}_i - {\p}_i {\t}^{jk} p_j \, {\l}_k, \qquad x^i
\mapsto x^i + {\t}^{ij}{\l}_j}This symmetry must be gauge fixed,
at the cost of introduction of a sequence of ghosts, anti-ghosts,
Lagrange multipliers and gauge conditions\ctn{cf}. It is not the
goal of these lectures to do so in full generality with regards to
various types of target space diffeomorphisms invariance one might
want to keep track of \ctn{bln}. Let us just mention that the
result of this gauge fixing procedure is the topological string
theory, which has some similarities both with the type A and type
B sigma models. The field content of this sigma model is the
following. It is convenient to think of $p_i$ and $x^i$ as of the
twisted super-fields, which simply mean that both $p_i$ and $x^i$
are differential forms on $\Sigma$ having components of all
degrees, $0,1,2$. The original Lagrangian had only $0$-th
component of $x^i$ and $1$st component of $p_i$. In addition,
there are auxiliary fields $\chi^i, H^i$, which are zero forms of
opposite statistics - $\chi$ of the fermionic while $H$ of the
bosonic one. The gauge fixing conditions restrict the components
of the superfields $p$ and $x$ to be: \eqn{gfx}{x^{i}_{(1)} =
\star {\rm d}\chi^i, \quad p_{i}^{(2)} = 0} We are discussing an
${\ap} \to 0$ limit of an open string theory. It means that the
worldsheet $\Sigma$ has a boundary, ${\p}{\Sigma}$. The fields $x,
p, \chi, H$ obey certain boundary conditions: $x$ obeys Neumann
boundary conditions, $H$ vanishes at ${\p}{\Sigma}$, i.e. it obeys
Dirichlet boundary conditions. In this sense the theory we are
studying is that of a D-brane wrapping the zero section ($H=0$)
$X$ of the tangent bundle $TX$ to the target space $X$. The rest
of boundary conditions is summarized in the formula for the
propagator below. The field $\chi^i$ is constant at the boundary
(fermionic Dirichlet condition).

The further treatment of the resulting theory is done in the
perturbation series expansion\ctn{k} in ${\t}$ around the
classical solution $x^{i}(z) = x^i = const$. We shall work on
$\Sigma$ being the upper half-plane ${\rm Im}z > 0$. The
propagator is most conveniently written in the superfield
language: \eqn{prpg}{\langle p_i (z) x^j (w) \rangle =
{\d}^{j}_{i} \, {\rm d} {\phi}(z, w), \quad {\phi}(z,w) =
{1\over{2\i}} \, {\rm log} {{(z-w)(z - {\wb})}\over{({\zb} -
{\wb})({\zb} - w)}} } where ${\rm d}$ is the deRham differential
on ${\Sigma} \times {\Sigma}$.

\subsubsection{Bulk observables}
The theory can be deformed by adding observables to the action.
Any  polyvector field ${\a}^{i_1 \ldots i_q}
{{\p}\over{{\p}x^{i_1}}} \wedge \ldots {{\p}\over{{\p}x^{i_q}}}
\in {\Lambda}^{q} TX$ on $X$ defines an observable in the theory:
\eqn{obs}{{\CO}_{\a} = {\a}^{i_1 \ldots i_q} p_{i_1} \ldots
p_{i_q}} which is again an inhomogeneous form on $\Sigma$. Its
degree $2$ component can be added to the action. In particular,
the ${\t}$-term in the original action corresponds to the
bi-vector ${1\over 2} {\t}^{ij} {\p}_i \wedge {\p}_j$. If we write
the action (before the gauge fixing) in the form: $$ \int p_i
\wedge {\rm d}x^i + {\CO}_{\a} $$ where ${\a}$ is a generic
polyvector field on $X$ then the only condition is that $[{\a},
{\a}] =0$ where $[,]$ is the Schoutens bracket on the polyvector
fields: \eqn{schbr}{[{\a}, {\b}] = \sum_i
{{{\p}{\a}}\over{{\p}x^i}} \wedge {{{\p}{\b}}\over{{\p}
({\p}_{i})}} \pm ({\a} \leftrightarrow {\b})} (the sign is
determined by the parity of the degrees of ${\a}, {\b}$, in such a
way that $[,]$ defines a super-Lie algebra).

\subsubsection{Boundary observables}

The correlation functions of interest are obtained by inserting
yet another observables at the boundary of $\Sigma$. These
correspond to the differential forms on $X$: $$ {\o}_{i_1 \ldots
i_q} dx^{i_1} \wedge \ldots \wedge dx^{i_q} \mapsto {\o}_{i_1
\ldots i_q} {\chi}^{i_1} \ldots {\chi}^{i_q}$$ Actually, one is
interested in computing string amplitudes. It means that the
positions of the vertex operators at the boundary must be
integrated over, up to the $SL_2({\IR})$ invariance (the gauge
fixed action still has (super)conformal invariance).

In particular, the three-point function on the disc (or two-point
function on the upper-half-plane, as a function of the boundary
condition at infinity, $x ({\infty}) = x$), produces the deformed
product on the functions: \eqn{crln}{\langle f(x(0)) g(x(1))
\left[ h(x)\chi\ldots \chi \right] (\infty) \rangle_{\t, \Sigma =
{\rm disc}} = \int_{X} f \star g \, h } for $f,g \in Fun(X)$, $h
\in {\Omega}^{{\rm dim}X}(X)$.

The $\star$-product defined by \rfn{crln} turns out to be
associative: $f \star (g \star l) = (f \star g) \star l$. This is
a consequence of a more general set of Ward identities obeyed by
the string amplitudes in the theory. For their description see
\ctn{k}, \ctn{cf}. The perturbative expression for the
$\star$-product following from \rfn{crln} turns out to be
non-covariant with respect to the changes of local coordinates in
$X$. This is a consequence of a certain anomaly in the theory. It
will be discussed in detail in \ctn{bln}.

From now on we go back to the case where $X$ is flat and ${\t}$ is
constant.

\vfill\eject\section{Instantons in noncommutative gauge theories}

{}We would like to  study the non-perturbative objects in
noncommutative gauge theory.

In this lecture specifically we shall be interested in four
dimensional instantons. They either appear as instantons
themselves in the Euclidean version of the four dimensional theory
(theory on Euclidean D3-brane), as solitonic particles in the
theory on D4-brane, i.e. in 4+1 dimensions, or as instanton
strings in the theory on D5-brane. They also show up as
``freckles'' in the gauge theory/sigma model
correspondence\ctn{freck}. We treat  only the bosonic fields, but
these could  be a part of a supersymmetric multiplet.

{\ndt}A D3-brane can be surrounded by other branes as well. For
example, in the Euclidean setup, a D-instanton could approach the
D3-brane. In fact, unless the D-instanton is dissolved inside
   the brane, the combined system breaks supersymmetry \ctn{witsei}. The
D3-D(-1) system can be rather simply described in terms of  a
noncommutative $U(1)$ gauge theory - the latter has instanton-like
solutions\ctn{neksch}.

{\ndt}More generally, one can have a stack of $k$ Euclidean
D3-branes with $N$D(-1)-branes inside. This configuration will be
described by charge $N$ instantons in $U(k)$ gauge theory.

Let us work in four Euclidean space-time dimensions, $i,j =
1,2,3,4$. As we said above, we shall look at the purely bosonic
Yang-Mills theory on the space-time ${\CA}_{\t}$ with the
coordinates functions $x^{i}$ obeying the Heisenberg commutation
relations: \eqn{he}{[ x^i, x^j ] = {\i} {\t}^{ij}} We assume that the
metric on the space-time is Euclidean: \eqn{mtr}{G_{ij} =
{\d}_{ij}}The action describing our gauge theory is given by:
\eqn{lgrn}{S = - {1\over{4g_{\rm YM}^2}} {\Tr} F \wedge * F \,
+ {{{\i}\vartheta}\over{4\pi}}  {\Tr} F \wedge F} where $g_{\rm YM}^2$ is
the Yang-Mills coupling constant, ${\vartheta}$ is that gauge theory theta angle, and \eqn{crvt}{F = {1\over 2}
F_{ij} dx^i \wedge dx^j, \quad F_{ij} = [ {\nabla}_i, {\nabla}_j
]} Here the covariant derivatives act $\nabla_i$ in the free
module $F_k$.

\subsection{Instantons}

\ndt {}The equations of motion following from \rfn{lgrn} are:
\eqn{eom}{\sum_{i} [{\nabla}_i,  F_{ij}] = 0} In general these
equations are as hard to solve as the equations of motion of the
ordinary non-abelian Yang-Mills theory. However, just like in the
commutative case\ctn{bpst}, there are special solutions, which are
simpler to analyze and which play a crucial role in the analysis
of the quantum theory. These are the so-called (anti-)instantons.
The (anti-)instantons solve the first order equation:
\eqn{asd}{F_{ij} = {\pm} {1\over{2}} \sum_{k,l} {\ve}_{ijkl}
F_{kl}}which imply \rfn{eom} by virtue of the Bianci identity:
$$[{\nabla}_i , F_{jk} ]  + \, {\rm cyclic} \, {\rm permutations}
\, = \, 0$$ These equations are easier to solve. The solutions are
classified by the instanton charge: \eqn{inch}{N = -
{1\over{8\pi^2}} {\Tr} F \wedge F}
and the action \rfn{lgrn} on such a solution is equal to:
\eqn{inac}{S_{N} = 2\pi {\i}{\tau}N, \qquad {\tau} = {{4\pi\i}\over{g_{\rm YM}^2}} + {{\vartheta}\over{2\pi}}}
Introduce the complex
coordinates: $z_1 = x_1 + i x_2 = x_{+}$, $z_2 = x_3 + i x_4$. The
instanton equations read: \eqn{insth}{F_{z_1 z_2} = 0, \, F_{z_1
{\zb}_1} + F_{z_2 {\zb}_2} = 0 } The first equation in \rfn{insth}
can be solved (locally) as follows: \eqn{ynga}{A_{{\zb}_a} =
{\xi}^{-1} {\pb}_{{\zb}_a} {\xi}, \quad A_{z_a} = - {\p}_{z_a}
{\xi} {\xi}^{-1}\ .} with $\xi$   a Hermitian matrix. Then the
second equation in \rfn{insth} becomes Yang's equation:
\eqn{ynge}{\sum_{a=1}^2 {\pb}_{z_a} \left( {\p}_{z_a} {\xi}^2
{\xi}^{-2} \right) = 0\ .} In the noncommutative case this ansatz
almost works globally (see below).

\subsection{ADHM construction}

In the commutative case all solutions to \rfn{asd} with the finite
action \rfn{lgrn} are obtained via the so-called
Atiyah-Drinfeld-Hitchin-Manin (ADHM) construction\ctn{adhm}.  If
we are concerned with the instantons in the $U(k)$ gauge group,
then the ADHM data consists of
\begin{enumerate}
\item the pair of the two complex vector spaces $V$ and $W$
of dimensions $N$ and $k$ respectively;

\item the operators: $B_{1}, B_{2} \in {\rm Hom} ( V, V)$, and
$I \in {\rm Hom} (W, V)$, $J \in {\rm Hom}(V, W)$;

\item the dual gauge group $G_{N} = U(N)$, which acts on the data above as
follows: \eqn{dgga}{B_{\a} \mapsto g^{-1} B_{\a} g, \,\, I \mapsto
g^{-1} I, \,\, J \mapsto J g}

\item Hyperk\"ahler quotient\ctn{rkh} with respect to the group \rfn{dgga}. It
means that one takes the set $X_{k,N} = {\m}_{r}^{-1}(0) \cap
{\m}_{c}^{-1}(0)$ of the common zeroes of the three moment maps:
\bea & {\m}_{r} = [B_{1} , B_{1}^{\dagger}] + [ B_{2},
B_{2}^{\dagger}] + II^{\dagger} - J^{\dagger}J \crt & {\m}_{c} =
[B_{1}, B_{2}] + IJ \\ & {\bar\m}_{c} = [ B_{2}^{\dagger},
B_{1}^{\dagger}] + J^{\dagger} I^{\dagger} \nonumber \label{mmnt}
\eea\\ and quotients it by the action of  $G_{N}$.
\end{enumerate}
The claim of ADHM is that the points in the space ${\CM}_{k,N} =
X_{k,N}^{\circ} /G_{N}$ parameterize the solutions to \rfn{asd}
(for ${\t}^{ij} =0$ ) up to the gauge transformations. Here
$X^{\circ}_{k,N} \subset X_{k,N}$ is the open dense subset of
$X_{k,N}$ which consists of the solutions to ${\vec \m} =0$ such
that their stabilizer in $G_{N}$ is trivial. The explicit formula
for the gauge field $A_{i}$ is also known. Define the Dirac-like
operator: \eqn{drc}{{\CD}^{+}  = \pmatrix{ -B_2 + z_2 & B_1 - z_1
& I \cr B_1^{\dagger} - {\zb}_1 & B_2^{\dagger} - {\zb}_2 &
J^{\dagger} \cr}: V \otimes {\IC}^2 \oplus W \to V \otimes
{\IC}^2} Here $z_1, z_2$ denote the complex coordinates on the
space-time: $$ z_1 = x_1 + i x_2,  \quad z_2 = x_3 + i x_4, \quad
{\zb}_1 = x_1 - ix_2, \quad {\zb}_2 = x_3 - ix_4 $$ The kernel of
the operator \rfn{drc} is the $x$-dependent vector space
${\CE}_{x} \subset  V \otimes {\IC}^2 {\oplus} W$. For generic
$x$, ${\CE}_{x}$ is isomorphic to $W$. Let us denote by ${\Psi} =
{\Psi}(x)$ this isomorphism. In plain words, ${\Psi}$ is the
fundamental solution to the equation: \eqn{fnde}{{\CD}^{+} {\Psi}
= 0, \qquad {\Psi}: W \to V \otimes {\IC}^2 {\oplus} W} If the
rank of ${\Psi}$ is $x$-independent (this property holds for
generic points in $\CM$), then one can normalize:
\eqn{nrml}{{\Psi}^{\dagger} {\Psi} = {\Id}_{k}}which fixes $\Psi$
uniquely up to an $x$-dependent $U(k)$ transformation ${\Psi}(x)
\mapsto {\Psi}(x) g (x)$, $g(x) \in U(k)$. Given $\Psi$ the
anti-self-dual gauge field is constructed simply as follows:
\eqn{asdg}{{\nabla}_{i} = {\p}_{i} + A_{i}, \qquad A_{i} =
{\Psi}^{\dagger}(x) {{\p}\over{{\p} x^{i}}} {\Psi} (x)} The space
of $(B_{0}, B_{1}, I, J)$ for which $\Psi (x)$ has maximal rank
for all $x$ is an open dense subset ${\CM}_{N,k} =
X^{\circ}_{N,k}/G_{N}$ in ${\CM}$. The rest of the points in
$X_{N,k}/G_{N}$ describes the so-called {\it point-like}
instantons. Namely, ${\Psi} (x)$ has maximal rank for all $x$ but
some finite set $\{ x_1, \ldots, x_l \}$, $l \leq k$. The
\rfn{nrml} holds for $x \neq x_i, \, i = 1, \ldots, l$, where the
left hand side of \rfn{nrml} simply vanishes.

{}The noncommutative deformation of the gauge theory leads to the
noncommutative deformation of the ADHM construction. It turns out
to be very simple yet surprising. The same data $V, W, B, I, J,
\ldots$ is used. The deformed ADHM equations are simply
\eqn{dfadhm}{{\m}_r = {\z}_r, \, {\m}_c = {\z}_c}where we have
introduced the following notations. The Poisson tensor ${\t}^{ij}$
entering the commutation relation $[x^i, x^j] = i{\t}^{ij}$ can be
decomposed into the self-dual and anti-self-dual parts
${\t}^{\pm}$. If we look at the commutation relations of the
complex coordinates $z_1, z_2, {\zb}_1, {\zb}_2$ then the
self-dual part of ${\t}$ enters the following commutators:
\eqn{cmtrs}{[z_1, z_2] = - {\z}_c \qquad [z_1, {\zb}_1] + [z_2,
{\zb}_2]  =  - {\z}_r} It turns out that as long as $\vert \z
\vert = {\z}_r^2 + {\z}_c {\bar\z}_c
>0$ one needs not to distinguish between ${\widetilde X}_{N,k}$
and ${\widetilde X}_{N,k}^{\circ}$, in other words the stabilizer
of any point in ${\widetilde X}_{N,k} = {\m}_{r}^{-1}(-{\z}_{r})
\cap {\m}_{c}^{-1}(-{\z}_{c})$ is trivial. Then the resolved
moduli space is ${\widetilde\CM}_{N,k} = {\widetilde
X}_{N,k}/G_{N}$.

By making an orthogonal rotation on the coordinates $x^{\m}$ we
can map the algebra ${\CA}_{\t}$ onto the sum of two copies of the
Heisenberg algebra. These two algebras can have different values
of ``Planck constants''. Their sum is the norm of the self-dual
part of ${\t}$, i.e. $\vert {\z} \vert$, and their difference is
the norm of anti-self-dual part of ${\t}$: \eqn{tcha}{[z_1,
{\zb}_1] = - {\z}_1, \, \quad [z_2, {\zb}_2] = -{\z}_2}where
${\z}_1 + {\z}_2 = \vert {\t}^{+} \vert$, ${\z}_1 - {\z}_2 = \vert
{\t}^{-}\vert$. By the additional reflection of the coordinates,
if necessary, one can make both ${\z}_1$ and ${\z}_2$ positive
(however, one should be careful, since if the odd number of
reflections was made, then the orientation of the space was
changed and the notions of the instantons and anti-instantons are
exchanged as well).

{}The next step in the ADHM construction was the definition of the
isomorphism ${\Psi}$ between the fixed vector space $W$ and the
fiber ${\CE}_{x}$ of the gauge bundle, defined as the kernel of
the operator ${\CD}^{+}$. In the noncommutative setup one can also
define the operator ${\CD}^{+}$ by the same formula \rfn{drc}. It
is a map between two free modules over ${\CA}_{\t}$:
\eqn{frm}{{\CD}_{x}^{+}: \left( V \otimes {\IC}^2 \oplus W \right)
\bigotimes {\CA}_{\t} \to \left( V \otimes {\IC}^2 \right)
\bigotimes {\CA}_{\t}} which commutes with the right action of
${\CA}_{\t}$ on the free modules. Clearly, $$ {\CE} = {\rm Ker}
{\CD}^{+}$$ is a right module over ${\CA}_{\t}$, for if ${\CD}^{+}
s = 0$, then ${\CD}^{+} (s \cdot a) = 0$, for any $a \in
{\CA}_{\t}$.

${\CE}$ is also a projective module, for the following reason.
Consider the operator ${\CD}^{+}{\CD}$. It is a map from the free
module $V \otimes {\IC}^2 \bigotimes {\CA}_{\t}$ to itself. Thanks
to \rfn{dfadhm} this map actually equals to ${\Delta} \otimes
{\Id}_{{\IC}^2}$ where ${\Delta}$ is the following map from the
free module $V \otimes {\CA}_{\t}$ to itself: \eqn{dlta}{\Delta =
(B_1 - z_1) (B_1^{\dagger} - {\zb}_1) + (B_2 - z_2 )(B_2^{\dagger}
- {\zb}_2) + II^{\dagger}} We claim that ${\Delta}$ has no kernel,
i.e. no solutions to the equation ${\Delta} \, v = 0$, $v \in V
\otimes {\CA}_{\t}$. Recall the Fock space representation ${\CH}$
of the algebra ${\CA}_{\t}$. The coordinates $z_{\a}, {\zb}_{\a}$,
obeying \rfn{tcha}, with ${\z}_1, {\z}_2 > 0$,  are represented as
follows: \eqn{rp}{z_1 = \sqrt{{\z}_1} \, c_1^{\dagger}, \, {\zb}_1
= \sqrt{{\z}_1} \, c_1 , \quad z_2 = \sqrt{{\z}_2} \,
c_2^{\dagger}, \, {\zb}_2 = \sqrt{{\z}_2} \, c_2}where $c_{1,2}$
are the annihilation operators and $c_{1,2}^{\dagger}$ are the
creation operators for the two-oscillators Fock space $${\CH} =
\bigoplus_{\scriptscriptstyle n_1, n_2 \geq 0} {\IC} \, \vert n_1,
n_2 \rangle$$ Let us assume the opposite, namely that there exists
a vector $v \in V \otimes {\CA}_{\t}$ such that ${\Delta} v = 0$.
Let us act by this vector on an arbitrary state $\vert n_1, n_2
\rangle$ in ${\CH}$. The result is the vector ${\n}_{\nb} \in V
\otimes {\CH}$ which must be annihilated by the operator
${\Delta}$, acting in $V \otimes {\CH}$ via \rfn{rp}. By taking
the Hermitian inner product of the equation ${\Delta} {\n}_{\nb}
=0$ with the conjugate vector ${\n}_{\nb}^{\dagger}$ we
immediately derive the following three equations: $$
(B_2^{\dagger} - {\zb}_2 ) {\n}_{\nb} \quad
 = \quad 0 $$ \eqn{triur}{(B_1^{\dagger} - {\zb}_1 ) {\n}_{\nb} \quad  =
\quad 0 } $$ I^{\dagger} {\n}_{\nb} \quad  = \quad 0 $$Using
\rfn{dfadhm} we can also represent ${\Delta}_x$ in the form:
\eqn{dltai}{{\Delta} = (B_1^{\dagger} - {\zb}_1) (B_1 - z_1) +
(B_2^{\dagger} - {\zb}_2) (B_2 - z_2) + J^{\dagger}J \ .} From
this representation another triple of equations follows: $$(B_2 -
z_2 ) {\n}_{\nb} \quad  = \quad 0 $$ \eqn{triura}{ (B_1 - z_1 )
{\n}_{\nb} \quad  = \quad 0} $$J {\n}_{\nb} \quad  = \quad 0 $$
Let us denote by $e_i$, $i=1, \ldots, N$ some orthonormal basis in
$V$. We can expand ${\n}_{\nb}$ in this basis as follows: $$
{\n}_{\nb} = \sum_{i=1}^{N} e_{i} \otimes v^i_{\nb}, \qquad
v^i_{\nb} \in {\CH}$$ The equations \rfn{triur},\rfn{triura}
imply: \eqn{repr}{(B_{\a})^i_j v^j_{\nb} = z_{\a} v^i_{\nb}, \quad
(B_{\a}^{\dagger})^{i}_j v^j_{\nb} = {\zb}_{\a} v^{i}_{\nb},
\qquad {\a} = 1,2} in other words the matrices $B_{\a},
B_{\a}^{\dagger}$ form a finite-dimensional representation of the
Heisenberg algebra which is impossible if either ${\z}_{1}$ or
${\z}_2 \neq 0$. Hence ${\n}_{\nb} =0$, for any ${\nb} = (n_1,
n_2)$ which implies that $v =0$. Notice that this argument
generalizes to the case where only one of ${\z}_{\a} \neq 0$.

Thus the Hermitian operator ${\Delta}$ is invertible. It allows to
prove the following theorem: each vector ${\psi}$ in the free
module $( V \otimes {\IC}^2 \oplus W) \otimes {\CA}_{\t}$ can be
decomposed as a sum of two orthogonal vectors: \eqn{dcmps}{ {\psi}
=
{\Psi}_{\psi} \oplus {\CD} {\chi}_{\psi}, \qquad {\CD}^{+}
{\Psi}_{\psi} = 0, \quad {\chi}_{\psi} \in (V \otimes {\IC}^2)
\otimes {\CA}_{\t}} where the orthogonality is understood in the
sense of the following ${\CA}_{\t}$-valued Hermitian product: $$
\langle \psi_1, \psi_2 \rangle = {\Tr}_{\scriptstyle V \otimes
{\IC}^2 \oplus W} \quad \left( {\psi}_1^{\dagger} {\psi}_2
\right)$$ The component ${\Psi}_{\psi}$ is annihilated by
${\CD}^{+}$, that is ${\Psi}_{\psi} \in {\CE}$. The image of
${\CD}$ is another right module over ${\CA}_{\t}$ (being the image
of the free module $(V \otimes {\IC}^2 ) \otimes {\CA}_{\t}$): $$
{\CE}^{\prime}  = {\CD} (V \otimes {\IC}^2 \otimes {\CA}_{\t})$$
and their sum is a free module: $$ {\CE} \oplus {\CE}^{\prime} =
{\CF} := \left( V \otimes {\IC}^2 \oplus W \right) \otimes
{\CA}_{\t}$$ hence ${\CE}$ is projective. It remains to give the
expressions for ${\Psi}_{\psi}, {\chi}_{\psi}$:
\eqn{pro}{{\chi}_{\psi} = {1\over {\CD}^{+}{\CD}} {\CD}^{+}
{\psi}, \quad {\Psi}_{\psi} = \Pi {\psi}, \quad {\Pi} = \left( 1 -
{\CD}{1\over {\CD}^{+}{\CD}} {\CD}^{+} \right)}The noncommutative
instanton is a connection in the module ${\CE}$ which is obtained
simply by projecting the trivial connection on the free module
${\CF}$ down to ${\CE}$. To get the covariant derivative of a
section $s \in {\CE}$ we view this section as a section of
${\CF}$, differentiate it using the ordinary derivatives on
${\CA}_{\t}$ and project the result down to ${\CE}$ again:
\eqn{cvrnt}{{\nabla} s = {\Pi} \, {\rm d} s} The curvature is
defined through ${\nabla}^2$, as usual: \eqn{crvtr}{{\nabla}
{\nabla} s = F \cdot s = {\rm d} {\Pi} \wedge {\rm d} {\Pi} \cdot
s} where we used the following relations: \eqn{prjctn}{{\Pi}^2 =
{\Pi}, \quad {\Pi} s = s} Let us now show explicitly that the
curvature \rfn{crvtr} is anti-self-dual, i.e.
\eqn{asdcnd}{[{\nabla}_i, {\nabla}_j ] + {\half} {\e}_{ijkl} [
{\nabla}_{k}, {\nabla}_{l}] = 0} First we prove the following
lemma: for any $s \in {\CE}$: \eqn{lemm}{{\rm d} {\Pi} \wedge {\rm
d} {\Pi} s = {\Pi} {\rm d} {\CD} {1\over{{\CD}^{+} {\CD}}} {\rm d}
{\CD}^{+} s} Indeed, $${\rm d} {\Pi} \wedge {\rm d} {\Pi} = {\rm
d} \left( {\CD} {1\over{{\CD}^{+} {\CD}}} {\CD}^{+} \right) \wedge
{\rm d} \left( {\CD} {1\over{{\CD}^{+} {\CD}}} {\CD}^{+} \right),
$$ $$ {\rm d} \left( {\CD} {1\over{{\CD}^{+} {\CD}}} {\CD}^{+}
\right)
 = {\Pi} {\rm d} {\CD} {1\over{{\CD}^{+} {\CD}}} {\CD}^{+}
 +  {\CD} {1\over{{\CD}^{+} {\CD}}} {\rm d} {\CD}^{+} {\Pi},
$$ $$ {\CD}^{+} {\Pi} = 0 $$ hence $$
 {\rm d} \left( {\CD} {1\over{{\CD}^{+}
{\CD}}} {\CD}^{+} \right)  \wedge  {\rm d} \left( {\CD}
{1\over{{\CD}^{+} {\CD}}} {\CD}^{+} \right) =
\qquad\qquad\qquad$$ $$ \qquad\qquad\qquad\qquad = {\Pi} {\rm d}
{\CD} {1\over{{\CD}^{+}{\CD}}} {\rm d} {\CD}^{+} {\Pi} +
{\CD} {1\over{{\CD}^{+}{\CD}}} {\rm d} {\CD}^{+}\, {\Pi}\,
{\rm d} {\CD} {1\over{{\CD}^{+}{\CD}}} {\CD}^{+} $$ and
the second term vanishes when acting on $s \in {\CE}$, while the
first term gives exactly what the equation \rfn{lemm} states.

Now we can compute the curvature more or less explicitly:
\eqn{crva}{F\cdot  s = 2 {\Pi} \pmatrix{{1\over{\Delta}} f_3 &
{1\over{\Delta}} f_+ & 0 \cr {1\over{\Delta}} f_{-} & -
{1\over{\Delta}} f_3  & 0 \cr 0 & 0 & 0 } \cdot s} where $f_{3},
f_{+}, f_{-}$ are the basic anti-self-dual two-forms on ${\bf
R}^4$: \eqn{sdf}{f_3 = {\half} \left( {\rm d} z_1 \wedge {\rm d}
{\zb}_1 - {\rm d} z_2 \wedge {\rm d} {\zb}_2 \right),\, f_{+}  =
{\rm d} z_1 \wedge {\rm d} {\zb}_2, \, f_{-}= {\rm d} {\zb}_1
\wedge {\rm d} z_2}Thus we have constructed the nonsingular
anti-self-dual gauge fields over ${\CA}_{\t}$. The interesting
feature of the construction is that it produces the non-trivial
modules over the algebra ${\CA}_{\t}$, which are projective for
any point in the moduli space. This feature is lacking in the
${\z} \to 0$, where it is spoiled by the point-like instantons.
This feature is also lacking if the deformed ADHM equations are
used for construction of gauge bundles directly over a commutative
space. In this case it turns out that one can construct a torsion
free sheaf over ${\IC}^2$, which sometimes can be identified with
a holomorphic bundle. However, generically this sheaf will not be
locally free. It can be made locally free by blowing up
sufficiently many points on ${\IC}^2$, thereby effectively
changing the topology of the space \ctn{branek}. The topology
change is rather mysterious if we recall that it is purely gauge
theory we are dealing with. However, in the supersymmetric case
this gauge theory is an ${\ap} \to 0$ limit of the theory on a
stack of Euclidean D3-branes. One could think that the topology
changes reflect the changes of topology of D3-branes embedded into
flat ambient space. This is indeed the case for monopole
solutions, e.g. \ctn{curtjuan,hashimoto,moriyama,mateos}. It is
not completely unimaginable possibility, but so far it has not
been justified (besides from the fact that the DBI solutions
\ctn{witsei,terashima} are ill-defined without a blowup of the
space). What makes this unlikely is the fact that the instanton
backgrounds have no worldvolume scalars turned on.

At any rate, the noncommutative instantons constructed above are
well-defined and nonsingular without any topology change.

Also note, that the formulae above define non-singular gauge
fields for any ${\t} \neq 0$. For ${\t}^{+} \neq 0$ these are
instantons, for ${\t}^{-} \neq 0$ they define anti-instantons (one
needs to perform an orthogonal change of coordinates which
reverses the orientation of ${\IR}^4$).

\subsubsection{The identificator $\Psi$}

In the \nct case one can also try to construct the identifying map
${\Psi}$. It is to be thought as of the homomorphism of the
modules over ${\CA}$: $$ {\Psi}: W \otimes {\CA}_{\t} \to {\CE}$$
The normalization \rfn{nrml}, if obeyed, would imply the unitary
isomorphism between the free module $W \otimes {\CA}_{\t}$ and
${\CE}$. We can write: ${\Pi} = {\Psi}{\Psi}^{\dagger}$ and the
elements $s$ of the module ${\CE}$ can be cast in the form:
\eqn{sec}{s = {\Psi}\cdot {\s}, \qquad {\s} \in W \otimes
{\CA}_{\t}} Then the covariant derivative can be written as:
\eqn{cvidnt}{{\nabla} s = {\Pi} d ({\Psi} \cdot{\s}) = {\Psi}
{\Psi}^{\dagger} d \left({\Psi}{\s} \right) = {\Psi} \left( d {\s}
+ A {\s} \right)} where \eqn{ggfld}{A = {\Psi}^{\dagger} d {\Psi}}
just like in the commutative case. For ${\rm Pf}({\t}) \neq 0$,
after having introduced the ``background independent''
operators\ctn{backsei}$D_i = i {\t}_{ij}x^{j} + A_{i}$, we write:
\eqn{bcind}{D_{i} = i {\Psi}^{\dagger} {\t}_{ij}x^{j} {\Psi}}

\section{Abelian instantons}

Let us describe the case of $U(1)$ instantons in detail, i.e. the
case $k=1$ in our notations above. Let us assume that ${\t}^{+}
\neq 0$. It is known, from\ctn{nakajima}, that for ${\z}_r
> 0, {\z}_c = 0$ the solutions to the deformed ADHM equations have
$J=0$. Let us denote by $V$ the complex Hermitian vector space of
dimensionality $N$, where $B_{\a}$, ${\a}= 1, 2$ act. Then $I$ is
identified with a vector in $V$. We can choose our units and
coordinates in such a way that ${\z}_r = 2, {\z}_c = 0$.

\subsection{Torsion free sheaves on ${\bf C}^2$}

Let us recall at this point the algebraic-geometric
interpretation\ctn{nakajima} of the space $V$ and the triple
$(B_{1}, B_{2}, I)$. The space ${\tilde X}_{N,1}$ parameterizes
the rank one torsion free sheaves on ${\IC}^2$. In the case of
${\bf C}^2$ these are identified with the ideals ${\CI}$ in the
algebra ${\CO} \approx {\IC}[z_1, z_2]$ of holomorphic functions
on ${\IC}^2$, such that $V = {\CO}/{\CI}$ has dimension $N$. An
ideal of the algebra ${\CO}$ is a subspace ${\CI} \subset {\CO}$,
which is invariant under the multiplication by the elements of
${\CO}$, i.e. if $g \in {\CI}$ then $f g \in {\CI}$ for any
${\CO}$.

An example of such an ideal is given by the space of functions of
the form: $$ f (z_1, z_2) = z_1^{N} g (z_1, z_2) + z_2 h(z_1, z_2)
$$The operators $B_{\a}$ are simply the operations of
multiplication of a function, representing an element of $V$ by
the coordinate function $z_{\a}$, and the vector $I$ is the image
in $V$ of the constant function $f = 1$. In the example above,
following\ctn{branek} we identify $V$ with ${\IC}[z_1]/z_1^{N}$,
the operator $B_2 = 0$, and in the basis $e_i = \sqrt{(i-1)!}
z_1^{N-i}$ the operator $B_1$ is represented by a Jordan-type
block: $B_1 e_i = \sqrt{2(i-1)} e_{i-1}$,  and $I = \sqrt{2N}
e_{N}$.

Conversely, given a triple $(B_1, B_2, I)$, such that the ADHM
equations are obeyed the ideal ${\CI}$ is reconstructed as
follows. The polynomial $f \in {\IC}[z_1, z_2]$ belongs to the
ideal, $f \in {\CI}$ if and only if $f(B_1, B_2) I = 0$. Then,
from the ADHM equations it follows that by acting on the vector
$I$ with polynomials in $B_1, B_2$ one generates the whole of $V$.
Hence ${\IC}[z_1, z_2]/{\CI} \approx V$ and has dimension $N$, as
required.

\subsection{Identificator $\Psi$ and projector $P$} Let us now
solve the equations for the identificator: ${\CD}^{\dagger}{\Psi}=
0, \, {\Psi}^{\dagger}{\Psi} = 1$. We decompose: \eqn{dciden}{\Psi
= \pmatrix{{\psi}_{+} \cr {\psi}_{-} \cr {\xi}}} where
${\psi}_{\pm} \in V \otimes {\CA}_{\t}$, ${\xi} \in {\CA}_{\t}$.
The normalization \rfn{nrml} is now:
\eqn{anrml}{{\psi}^{\dagger}_{+}{\psi}_{+} +
{\psi}^{\dagger}_{-}{\psi}_{-} + {\xi}^{\dagger}{\xi} = 1}It is
convenient to work with rescaled matrices $B$:
 $B_{\a} = \sqrt{\z_{\a}} {\b}_{\a}$, ${\a}= 1,2$.
The equation ${\CD}^{\dagger}{\Psi}=0$ is solved by the
substitution: \eqn{ansatz}{{\psi}_{+}=  - \sqrt{{\z}_2} (
{\b}_2^{\dagger} - c_{2}) v, \quad {\psi}_{-} = \sqrt{{\z}_1} (
{\b}_{1}^{\dagger} - c_1 ) v} provided \eqn{master}{{\hat\Delta} v
+ I {\xi} = 0, \qquad {\hat\Delta} = \sum_{\a} {\z}_{\a}
({\b}_{\a} - c^{\dagger}_{\a})({\b}_{\a}^{\dagger} -
c_{\a})}Fredholm's alternative states that the solution ${\xi}$ of
\rfn{master} must have the property, that for any  ${\n} \in
{\CH}, {\chi} \in V$, such that \eqn{babush}{{\hat\Delta} ({\n}
\otimes {\chi}) = 0,} the equation \eqn{fred}{\left(
{\n}^{\dagger}\otimes {\chi}^{\dagger} \right) I {\xi} = 0}holds.
It is easy to describe the space of all ${\n} \otimes {\chi}$
obeying \rfn{babush}: it is spanned by the vectors:
\eqn{spnd}{e^{\sum {\b}_{\a}^{\dagger} c_{\a}^{\dagger}} \vert
0,0\rangle \otimes e_i, \quad i = 1, \ldots, N} where $e_i$ is any
basis in $V$. Let us introduce a Hermitian operator $G$ in $V$:
\eqn{gop}{G = \langle 0,0 \vert e^{\sum {\b}_{\a}c_{\a}}
II^{\dagger} e^{\sum {\b}_{\a}^{\dagger} c_{\a}^{\dagger}} \vert
0,0\rangle} It is positive definite, which follows from the
representation: $$ G = \langle 0,0 \vert e^{\sum {\b}_{\a}c_{\a}}
\sum {\z}_{\a} ({\b}_{\a}^{\dagger} - c_{\a})({\b}_{\a} -
c_{\a}^{\dagger}) e^{\sum {\b}_{\a}^{\dagger} c_{\a}^{\dagger}}
\vert 0,0\rangle $$ and the fact that ${\b}_{\a} -
c_{\a}^{\dagger}$ has no kernel in ${\CH} \otimes V$. Then define
an element of the algebra ${\CA}_{\t}$ \eqn{prj}{P = I^{\dagger}
e^{\sum {\b}_{\a}^{\dagger} c_{\a}^{\dagger}} \vert 0,0 \rangle
G^{-1} \langle 0,0 \vert e^{\sum {\b}_{\a} c_{\a}} I} which obeys
$P^2 = P$, i.e. it is a projector. Moreover, it is a projection
onto an $N$-dimensional subspace in ${\CH}$, isomorphic to $V$.

\subsection{Dual gauge invariance}

The normalization condition \rfn{nrml} is invariant under the
action of the dual gauge group $G_{N} \approx U(N)$ on $B_{\a},
I$. However, the projector $P$ is invariant under the action of
larger group - the complexification
$G_{N}^{\scriptscriptstyle{\IC}} \approx GL_{N}({\IC})$:
\eqn{dualc}{(B_{\a}, I) \mapsto (g^{-1}B_{\a}g, g^{-1}I), \quad
(B_{\a}^{\dagger}, I^{\dagger}) \mapsto
(g^{\dagger}B_{\a}g^{\dagger, -1}, I^{\dagger}g^{\dagger, -1})}
This makes the computations of $P$ possible even when the solution
to the ${\m}_{r} = {\z}_{r}$ part of the ADHM equations is not
known. The moduli space ${\widetilde\CM}_{N,k}$ can be described
both in terms of the hyperkahler reduction as above, or in terms
of the quotient of the space of stable points $Y_{N,k}^{s} \subset
{\m}_{c}^{-1}(0)$ by the action of
$G_{N}^{\scriptscriptstyle{\IC}}$ (see \ctn{donaldson,nakajima}
for related discussions). The stable points $(B_1, B_2, I)$ are
the ones where $B_1$ and $B_2$ commute, and generate all of $V$ by
acting on $I$: ${\IC}[ B_1, B_2]\, I = V$, i.e. precisely those
triples which correspond to the codimension $N$ ideals in
${\IC}[z_1, z_2]$.

\subsection{Instanton gauge field} Clearly, $P$ annihilates
${\xi}$, thanks to \rfn{fred}. Let $S$ be an operator in ${\CH}$
which obeys the following relations: \eqn{iden}{SS^{\dagger} = 1,
\quad S^{\dagger}S = 1- P} The existence  of $S$ is merely a
reflection of the fact that as Hilbert spaces ${\CH}_{\CI} \approx
{\CH}$. So it just amounts to re-labeling the orthonormal bases in
${\CH}_{\CI}$ and ${\CH}$ to construct $S$. The operators $S,
S^{\dagger}$ were introduced in the \nct gauge theory context in
\ctn{grossneki} and played a prominent role in the constructions
of various solutions of \nct gauge theories, e.g. in
\ctn{grossneki,amgs,fkiii,klh,grossnekii,fkiv}.

Now, ${\hat\Delta}$ restricted at the subspace
$S^{\dagger}{\CH}\otimes I \subset {\CH} \otimes V$, is
invertible. We can now solve \rfn{master} as follows:
\eqn{slmaster}{{\xi} = {\Lambda}^{-\half} S^{\dagger}, v = -
{1\over{\hat\Delta}}I {\xi}} where \eqn{mslmbd}{{\Lambda} = 1 +
I^{\dagger}{1\over{\hat\Delta}}I} ${\Lambda}$ is not an element of
${\CA}_{\t}$, but ${\Lambda}^{-1}$ and ${\Lambda}S^{\dagger}$ are.
Finally, the gauge fields can be written as: \eqn{gage}{D_{\a} =
\sqrt{1\over{{\z}_{\a}}} S {\Lambda}^{-\half} c_{\a}
{\Lambda}^{\half} S^{\dagger}, \qquad {\bar D}_{\bar \a} = -
\sqrt{1\over{{\z}_{\a}}} S {\Lambda}^{\half} c_{\a}^{\dagger}
{\Lambda}^{-\half} S^{\dagger}}{\small If ${\z}_1 {\z}_2 = 0$ then
the formula \rfn{gage} must be modified in an interesting way. We
leave this as an exercise.} Notice that if $S^{\dagger}S$ was
equal to $1$ then the expressions \rfn{gage} had the Yang form
\rfn{ynga}.
\subsection{Ideal meaning of $P$} We can explain the meaning of
$P$ in an invariant fashion, following\ctn{fki}. Consider the
ideal ${\CI}$ in ${\IC}[z_1, z_2]$, corresponding to the triple
$(B_1, B_2, I)$ as explained above. Any polynomial $f {\in} {\CI}$
defines a vector $f ( \sqrt{{\z}_1} c_1^{\dagger}, \sqrt{{\z}_2}
c_2^{\dagger} ) \vert 0,0 \rangle$ and their totality span a
subspace ${\CH}_{\CI} \subset {\CH}$ of codimension $N$. The
operator $P$ is simply an orthogonal projection onto the
complement to ${\CH}_{\CI}$. The fact ${\CI}$ is an ideal in
${\IC}[z_1, z_2]$ implies that $c_{\a}^{\dagger} ({\CH}_{\CI})
\subset {\CH}_{\CI}$, hence: $$ c_{\a}^{\dagger} S^{\dagger}
{\eta} = S^{\dagger} {\eta}^{\prime} $$ for any ${\eta} \in
{\CA}_{\t}$, and also ${\Lambda}^{-\half} S^{\dagger} =
S^{\dagger} {\eta}^{\prime\prime}$ for some ${\eta}^{\prime},
{\eta}^{\prime\prime} \in {\CA}_{\t}$.

Notice that the expressions \rfn{gage} are well-defined. For
example, the ${\bar D}_{\bar \a}$ component contains a dangerous
piece ${\Lambda}^{\half} c_{\a}^{\dagger}\ldots$ in it. However,
in view of the previous remarks it is multiplied by $S^{\dagger}$
from the right and therefore well-defined indeed.

\subsection{Charge one instanton} In this  case: $I = \sqrt{2}$,
one can take $B_{\a} =0$, ${\hat\Delta} = \sum {\z}_{\a} n_{\a}$,
and in addition we shall assume that ${\z}_1 {\z}_2 \neq 0$. $$
{\Lambda} = {{M + 2}\over{M}}$$ $M = \sum_{\a} {\z}_{\a} n_{\a},
\, \sum_{\a} {\z}_{\a} = 2$. Let us introduce the notation $N =
n_1 + n_2$. For the pair ${\nb} = (n_1, n_2)$ let $\rho_{\nb} =
{\half}N(N-1) + n_1$. The map ${\nb} \leftrightarrow \rho_{\nb}$
is one-to-one. Let $S^{\dagger} \vert {\rho}_{\nb} \rangle = \vert
{\rho}_{\nb} + 1 \rangle$. Clearly, $SS^{\dagger} = 1, \,
S^{\dagger}S = 1  - \vert 0,0\rangle\langle 0,0 \vert$.

The formulae \rfn{gage} are explicitly non-singular. Let us
demonstrate the anti-self-duality of the gauge field \rfn{gage} in
this case. $$ \sum_{\a} D_{\a}{\bar D}_{\bar\a} = - S
{1\over{{\z}_{\a}}} (n_{\a} + 1) {M\over{M+2}}
{{M+2+{\z}_{\a}}\over{M+{\z}_{\a}}} S^{\dagger} $$ $$ \sum_{\a}
{\bar D}_{\bar\a}D_{\a} = S {1\over{{\z}_{\a}}} n_{\a}
{{M-{\z}_{\a}}\over{M+2-{\z}_{\a}}} {{M+2}\over{M}} S^{\dagger}$$
A simple calculation shows: \eqn{crvaa}{\sum_{\a} [ D_{\a}, {\bar
D}_{\bar\a} ] = - {2\over{{\z}_1{\z}_2}} = - \left(
{1\over{{\z}_1}} + {1\over{{\z}_2}} \right), \qquad [D_{\a},
D_{\b}] = 0,} hence \eqn{asdcrva}{\sum_{\a} F_{\a{\bar\a}} = 0} as
$$ i\sum_{\a} {\t}_{\a\bar\a} = {1\over{{\z}_1}} +
{1\over{{\z}_2}} $$ This is a generalization of the charge one
instanton constructed in \ctn{neksch}, written in the explicitly
non-singular gauge. The explicit expressions for higher charge
instantons are harder to write, since the solution of the deformed
ADHM equations is not known in full generality. However, in the
case of the so-called ``elongated'' instantons\ctn{branek} the
solution can be written down rather explicitly, with the help of
Charlier polynomials. In the $U(2)$ case it is worth trying to
apply the results of\ctn{rational} for the studies of the
instantons of charges $\leq 3$.

\subsection{Remark on gauges}

The gauge which was chosen in the examples considered in
\ctn{neksch} and subsequently adopted in \ctn{fki,gms} had ${\xi}
= {\xi}^{\dagger}$. It was shown in \ctn{fki} that this gauge does
not actually lead to the canonically normalized identificator
${\Psi}$: one had ${\Psi}^{\dagger}{\Psi} = 1 - P$. This gauge is
in some sense an analogue of 't Hooft singular gauge for
commutative instantons: it leads to singular formulae, if the
gauge field is considered to be well-defined globally over
${\CA}_{\t}$ (similar observations were made previously in
\ctn{fkii}). However, as we showed above, there are gauges in
which the gauge field is globally well-defined, non-singular, and
anti-self-dual. They simply have ${\xi} \neq {\xi}^{\dagger}$.

\vfill\eject\section{Monopoles in noncommutative gauge theories}

\subsection{Realizations of monopoles via D-branes}

Another interesting BPS configuration of D-branes is that of a
D-string that ends on a D3-brane. The endpoint of the D-string is
a magnetic charge for the gauge field on the D3-brane. In the
commutative case, in the absence of the $B$-field, the D-string is
a straight line, orthogonal to the D3-brane. It projects onto the
D3-brane in the form of a singular source, located at the point
where the D-string touches the  D3-brane. From the point of view
of the D3-brane this is a Dirac monopole, with   energy density
that diverges at the origin.

The situation changes drastically when the $B$-field is turned on.
One can trade a constant background $B$-field with spatial
components for a constant background magnetic field. The latter
pulls the magnetic monopoles with the constant force. As a
consequence, the D-string bends \ctn{hashimoto}, in such a way
that its tension compensates the magnetic force. It projects to
the D3-brane as a half-line with finite tension. It is quite
fascinating to see that the shadow of this string is seen by the
noncommutative gauge theory. The $U(1)$ noncommutative gauge
theory with adjoint Higgs field has a monopole solution
\ctn{grossnek}, which is everywhere non-singular, and whose energy
density is peaked along a half-line, making up a semi-infinite
string. The non-singularity of the solution is non-perturbative in
${\t}$ and cannot be seen by the expansion in ${\t}$ around the
Dirac monopole \ctn{genmnp}.

This analytic solution extends to the case of $U(2)$
noncommutative gauge theory. In this case one finds strings of
finite extent, according to the brane picture \ctn{hashimoto},
which has the D-string suspended between two D3-branes separated
by a finite distance. The solitons in $U(1)$ theory were localized
in the noncommutative directions, but generically occupied all of
the commutative space, corresponding to (semi)infinite
D(p-2)-branes, immersed in  a Dp-brane, or piercing it. In the
case of several Dp-branes we shall describe solitons which,
although they have finite extent in the commutative directions,
are nevertheless localized and look like codimension three objects
when viewed from far away. The simplest such object is the
monopole in the \nct $U(2)$ gauge theory, i.e. the theory on a
stack of two separated D3-branes  in the Seiberg-Witten limit
\ctn{witsei}.

The fact that all the fields involved are non-singular, and that
the solution is in fact a solution to the noncommutative version
of the Bogomolny equations everywhere, shows that the string in
the monopole solution is an intrinsic object of the gauge theory.
As such, one could expect that the noncommutative gauge theory
 describes strings as well.  In fact, the
\nct gauge theory has solutions, describing infinite magnetic flux
strings, whose fluctuations match with those of D-strings, located
anywhere in the ten dimensional space around the D3-branes
\ctn{grossnekii}.

\subsection{Monopole equations}

{}If we look for the solutions to \rfn{asd}, that are invariant
under translations in the $4$'th direction then we will find the
monopoles of the gauge theory with an adjoint scalar Higgs field,
where the role of the Higgs field is played by the component $A_4$
of the gauge field. The equations \rfn{asd} in this case are
called the  Bogomolny equations, and they can be analyzed in the
commutative case via Nahm's ansatz \ctn{nahm}.

{}For the $x_4$-independent field configurations the action
\rfn{glagr} produces the energy functional for the coupled
gauge-adjoint Higgs system: \eqn{enrg}{{\CE} = {{2\pi \t}
\over{4g_{\rm YM}^2}} \int {\rm d}x_3 \, \sqrt{{\rm det}G} {\Tr}
\left( - G^{ii^{\prime}} G^{jj^{\prime}} F_{ij} \star
F_{i^{\prime}j^{\prime}} + 2 \, G^{ij} {\nabla}_i {\Phi} \star
{\nabla}_j {\Phi} \right)} where for the sake of generality we
have again introduced a constant metric $G_{ij}$. As before, the
factor $2{\pi}{\t}$ comes from relating the integral over $x^1,
x^2$ to the trace over the Fock space which replaces the
integration over the \nct part of the three dimensional space. The
trace also includes the summation over the color indices, if they
are present (for several D3-branes). In terms of the three
dimensional gauge fields and the adjoint Higgs the Bogomolny
equations have the form: \eqn{bgmlni}{{\nabla}_i {\Phi} \, = \,
\pm B_i, \quad i = 1,2,3\ .} where \rfn{mgnf} in the case of
generic metric $G$ has the form: $$ B_i = {{\i}\over 2} {\ve}_{ijk}
 G^{jj^{\prime}} G^{kk^{\prime}} \sqrt{G} \,\, F_{j^{\prime}k^{\prime}}$$

As in the ordinary, commutative case, one can rewrite \rfn{enrg}
as a sum of a total square and a total derivative:
\eqn{splt}{{\CE} = \matrix{& {{\pi\t}\over{g_{\rm YM}^2}} \int
{\rm d} x_3 \sqrt{G} G^{ij} {\Tr} \left( {\nabla}_{i} {\Phi} \pm
B_i \right) \star \left( {\nabla}_j \Phi \pm B_j \right) \cr & \mp
{\p}_j \left[ \sqrt{G} G^{ij} {\Tr} \left( B_i \star {\Phi} +
{\Phi} \star B_i \right)\right]}} The total derivative term
depends only on the boundary conditions. So, to minimize the
energy given boundary conditions we should solve the {\it
Bogomolny} equations \rfn{bgmlni}.

\subsection{Nahm's construction}

\subsubsection{Ordinary monopoles}

\ndt{}To begin with, we review the techniques which  facilitate
the solution of the ordinary Bogomolny equations:
\eqn{bgmln}{{\nabla}_i {\Phi} + B_i = 0, \quad i = 1,2,3}They are
supplemented with the boundary condition that
  at the  spatial infinity the norm of the Higgs field
approaches a constant, corresponding to the Higgs vacuum.
  In the case of $SU(2)$ this means   that locally on the
two-sphere at infinity: \eqn{asmt}{{\phi} (x) \sim {\rm diag}
\left( {a\over 2}, - {a\over 2} \right)\ .} The solutions are
classified by the magnetic charge $N$. By virtue of the equation
\rfn{bgmln} the monopole charge can be expressed as the winding
number which counts  how many times the two-sphere ${\bf
S}^2_{\infty}$ at infinity is mapped to the coset space
$SU(2)/U(1) \approx {\bf S}^2$ of the abelian subgroups of the
gauge group. We shall present a general discussion of the charge
$N$ monopoles in the gauge group $U(k)$, where $k$ will be either
$1$ or $2$.

The approach to the solution of \rfn{bgmln} is via the
modification of the ADHM construction. After all, \rfn{bgmln} are
also instanton equations, with different asymptotic conditions on
the gauge fields.  The appropriate modification of the ADHM
construction was found by Nahm. Nahm \ctn{nahm} constructs
solutions to the monopole equations as follows. Consider the
matrix differential operator on the interval $I$ with the
coordinate $z$: \eqn{tmatr}{-{\i} {\Delta} =
  {\p}_{z} +  {\CT}_i {\s}_i, }
where \eqn{tmatri}{{\CT}_i = T_i(z) + x_i \ .} $x_i$ are the
coordinates in the physical space ${\bf R}^3$, and the $N \times
N$ matrices $T_i(z) = T_i^{\dagger} (z)$ obey Nahm's equations:
\eqn{nms}{{\p}_{z} T_i = {{\i}\over 2} {\ve}_{ijk} [T_j , T_k]\ ,
}with certain boundary conditions.

The range of the coordinate $z$ depends on the specifics of the
problem we are to address. For the $SU(2)$ monopoles with the
asymptotics\rfn{asmt} we take $I = ( -a/2, a/2)$ where $a$ is
given in \rfn{asmt}. For the $U(1)$ Dirac monopoles we would take
$I = (-\infty, 0)$. At $z$ approaches the boundary of $I$, $z \to
z_0$ we require that : \eqn{asmtt}{T_i \sim {t_i \over z - z_{0}}
+ {\rm reg.}, \quad [ t_i, t_j] = {\i} {\ve}_{ijk} t_k\ ,} i.e. the
residues $t_i$ must form a $N$-dimensional representation of
$SU(2)$ (irreducible if the solution is to be non-singular).

\noindent Then one looks for the fundamental solution to the
equation: \eqn{drci}{-i {\Delta}^{\dagger} {\Psi} (z) =  {\p}_z
{\Psi} -  {\CT}_i {\s}_i {\Psi} = 0\ ,} where $${\Psi} =
\pmatrix{{\Psi}_{+} \cr {\Psi}_{-}},$$ and ${\Psi}_{\pm}$ are $N
\times k$ matrices, which must be finite at ${\p}I$ and normalized
so that: \eqn{nrmlzi}{\int_{I} dz \,\, {\Psi}^{\dagger} {\Psi} =
{\bf 1}_{k \times k}\ .}

\ndt Then: \eqn{sltn}{A_i  \quad = \quad \int_{I} dz \,
{\Psi}^{\dagger} {\p}_i {\Psi}, } $$ {\Phi} =  \int_{I} dz \, z \,
{\Psi}^{\dagger} {\Psi}\ . $$

\noindent Notice that for $k=2$  the interval $I$ could be $(a_1,
a_2)$ instead of $(-a/2, +a/2)$. The only formula that is not
invariant under shifts of $z$ is the expression \rfn{sltn} for
$\phi$. By shifting $\Phi$ by a scalar $(a_1 + a_2)/2$ we can make
it traceless  and map $I$ back to the form we used above.

\subsubsection{Abelian ordinary monopoles}

In the case $k = 1$ Nahm's equations describe Dirac monopoles.
Take $I = ( - \infty, 0)$. The equation \rfn{bgmln} becomes simply
the condition that the abelian monopole has a magnetic potential
$\Phi$, which must be harmonic. Let us find this harmonic
function. The matrices $T_i$ can be taken to have the following
form: \eqn{sutwo}{T_i (z) = {t_i \over z }, \quad [t_i , t_j] = {\i}
{\ve}_{ijk} t_k \ ,} where $t_i$ form an irreducible spin $j$
representation of $SU(2)$. Let $V \approx {\IC}^{N}, \,  N =
2j+1$, be the space of this representation. The matrices
$\Psi_{\pm}$ are now $V$-valued. By an $SU(2)$ rotation we can
bring the three-vector $x_i$ to the form $(0,0,r)$, i.e. $x_1 =
x_2 = 0, x_3 > 0$. Then in this basis:\eqn{ps}{{\Psi}_{-} = 0,
\quad {\Psi}_{+} =  {\sqrt{2r} \over \sqrt{(N-1)!}} (2rz)^{j}
e^{rz} \vert j \rangle \ ,}where $\vert j \rangle \in V$ is the
highest spin state in $V$.  From this we get the familiar formula
for the singular Higgs field \eqn{sing}{ \Phi = - {N\over 2r}\ .}
corresponding to $N$ Dirac monopoles sitting on top of each other.
\subsubsection{Nonabelian ordinary monopoles.}
\ndt We now take $k = 2, N = 1$. Let $H = {\IC}^{k}$ be the
Chan-Paton space, the fundamental representation of the gauge
group. Let $e_{0}, e_{1}$ be the orthonormal basis in $H$. Again,
for $N=1$ the analysis of the equation \rfn{nms} is simple: $T_i =
0$. We can take $a_{\pm} = {\pm}{a\over 2}$ and \eqn{fndmi}{{\Psi}
= \pmatrix{ \left( {\p}_z + x_3 \right) v \cr \left( x_1 + {\i} x_2
\right) v }, \quad {\p}_z^2 v = r^2 v, \quad r^2 = \sum_i x_i^2 \
.}The condition that $\Psi$ is finite at both ends of the interval
allows for  two solutions of \rfn{drc} in the form of \rfn{fndmi}:
$$ v = e^{\pm r z}, $$ which after imposing the normalization
condition,\rfn{nrmcm}, leads to: $$ {\Psi} = {1\over \sqrt{2{\rm
sinh} (ra)}}\left[ e^{rz} \pmatrix{\sqrt{r + x_3}\cr {x_{+} \over
\sqrt{r + x_3}}\cr} \otimes e_{0} + e^{-rz} \pmatrix{ - \sqrt{r -
x_3}\cr {x_{-} \over \sqrt{r - x_3}}\cr} \otimes e_{1}\right], $$
where $x_{\pm} = x_1 {\pm} {\i} x_2$.

In particular, $$ {\Phi} = {1\over 2}\left( {a\over  {\rm tanh}
(ra)} - {1\over r} \right) {\s}_{3}. $$ This is famous 't
Hooft-Polyakov monopole of the higgsed $SU(2)$ gauge theory.

\subsubsection{Nahm's equations from the D-string point of view}

The meaning of the Nahm's equations becomes clearer in the D-brane
realization of   gauge theory and  the D-string construction of
monopoles. The endpoint of a fundamental string touching a
D3-brane looks like an electric charge for the $U(1)$ gauge field
on the brane. By S-duality, a D-string touching a D3-brane creates
a magnetic monopole. If one starts with two parallel D3-branes,
seperated by
  distance $a$ between them, one  is studying the $U(2)$
gauge theory, Higgsed down to $U(1) \times U(1)$, where the vev of
the Higgs field is $$ {\Phi} = \pmatrix{ a_1 & 0 \cr 0 & a_2 \cr}
$$ One can suspend a D-string between these two D3-branes, or a
collection of $k$ parallel D-strings. These would correspond to a
charge $k$ magnetic monopole in the Higgsed $U(2)$ theory. The BPS
configurations of these D-strings are described the corresponding
self-duality equations in the 1+1 dimensional $U(k)$ gauge theory
on the worldsheet of these strings \ctn{manuel}, $z$ being the
spatial coordinate along the D-string. The equations \rfn{nms} are
exactly these BPS equations. The presence of the D3-branes is
reflected in the boundary conditions \rfn{asmtt}. The matrices
$T_i$ correspond to the ``matrix'' transverse coordinates $X^i$,
$i=1,2,3$ to the D-strings, which lie within D3-branes. One can
also consider the BPS configurations of semi-infinite D-strings,
in which case the parameter $z$ lives on a half-line. For example
a collection of $N$ D-strings ending on the D3-brane forms the
so-called BIon\ctn{bion}, described by the solution\rfn{sutwo}.

\subsubsection{Old point of view}

The old-fashioned point of view at the equations
\rfn{drci},\rfn{fnde} is that they are the equations obeyed by the
kernel of the family of the Dirac operators in the background of
the gauge/Higgs fields obeying self-duality
condition\ctn{nahm,cg}. This interpretation also holds in the \nct
case\ctn{neksch,neksau}.

\subsubsection{Noncommutative monopole equations}

Now let us study the solutions to the Bogomolny equations for a
gauge theory on a noncommutative three dimensional space. As
before, we assume the Poisson structure ($\t$) which deforms the
multiplication of the functions to be constant. Then there is
essentially a unique choice of  coordinate functions $x_1, x_2,
x_3$ such that the commutation relations between them are as
follows (cf. \rfn{ncm}): $${[ x_1, x_2 ] = - {\i}{\t}, \, {\t}
> 0\qquad
[x_1 , x_3 ] = [x_2, x_3 ] = 0 \ . }$$ This  algebra defines
noncommutative ${\bf R}^3$, which  we still denote   by
${\CA}_{\t}$. Introduce the creation and annihilation operators
$c, c^{\dagger}$: \eqn{osci}{c = {1\over{\sqrt{2\t}}} \left( x_1 -
{\i} x_2 \right), \quad c^{\dagger} = {1\over{\sqrt{2\t}}} \left( x_1
+ {\i} x_2 \right)\ , } that  obey $$ [c , \, c^{\dagger} ] = 1.$$ We
wish to solve Bogomolny equations\rfn{bgmln}, which can also be
written as: \eqn{bgmlnii}{[D_i, {\Phi}] = {{\i}\over{2}} {\ve}_{ijk}
[D_j, D_k] - {\d}_{i3} {1\over {\t}} \ , } where ${\Phi}$ and
$D_i$, $i=1,2,3$ are the $x_3$-dependent operators in ${\CH}
\otimes H$. Now the relation between $D_i$ and $A_i$ is as
follows: \eqn{bckg}{D_3 = {\p}_3 + A_3, \qquad D_{\a} = {\i}
{\t}^{-1} {\ve}_{\a\b} x^{\b} + A_{\a}, \quad {\a},{\b} = 1,2}

\subsubsection{Noncommutative Nahm equations}

\ndt We proceed {\it a la} \ctn{neksch} by repeating the procedure
that worked in the ADHM instanton case, namely we relax the
condition that $x_i$'s commute but insist on the equation
\rfn{nms} with $T_i$ replaced by the relevant matrices ${\CT}_i =
T_i + x_i$. Then the equation \rfn{nms} on $T_i$ is modified:
\eqn{mdfnms}{{\p}_z T_i = {{\i}\over 2} {\ve}_{ijk} [T_j , T_k ] +
{\d}_{i3} {\t}\ .} It is obvious that, given a solution $T_i(z)$
of the original Nahm equations\rfn{nms}, it is easy to produce a
solution of the noncommutative ones: \eqn{mdf}{T_{i}(z)^{\rm nc} =
T_{i}(z) + {\t} z {\d}_{i3}\ . }From this it follows that, unlike
the case of instanton moduli space, the monopole moduli space {\it
does not change under noncommutative deformation}.

Notice the similarity of \rfn{mdfnms} and \rfn{bgmlnii}, which
becomes even more striking if we take into account that the
spectral meaning of the coordinate $z$ as the eigenvalue of the
operator ${\Phi}$. This similarity is explained in the framework
of {\it noncommutative reciprocity}\ctn{neksau}, generalizing the
commutative reciprocity\ctn{cg,nakajimamon}

{\ndt}The deformation\rfn{mdfnms} is exactly what one gets by
looking at the D-strings suspended between the
D3-branes\ctn{bak,moriyama} (or a semi-infinite D-string with one
end on a D3-brane) in the presence of a $B$-field. One gets the
deformation: \eqn{dfrm}{ [X^i, X^j] \to [X^i, X^j] - {\i}{\t}^{ij} =
[T_i, T_j] - {1\over 2} {\t} {\ve}_{ij3}} The reason why
${\t}^{ij}$, instead of $B_{ij}$, appears on the right hand side
of \rfn{dfrm} is rather simple. By applying T-duality in the
directions $x_1, x_2, x_3$ we could map the D-string into the
D4-brane. The matrices $X^1, X^2, X^3$ become the components
$A_{\hat 1}, A_{\hat 2}, A_{\hat 3}$ of the gauge field on the
D4-brane worldvolume, and the $B$-field would couple to these
gauge fields via the standard coupling $F_{{\hat i}{\hat j}} -
{\hat B}_{{\hat i}{\hat j}}$, where ${\hat B}_{{\hat i}{\hat j}}$
 is the T-dualized $B$-field. It remains to observe that
${\hat B}_{\hat i \hat j} = {\t}^{ij}$, since the T-dualized
indices ${\hat i}$ label the coordinates on the space, dual to
that of $x_i$'s.

\subsection{Solving Nahm's equations for noncommutative monopoles}

\ndt{}To solve \rfn{mdfnms}  we imitate the approach for the
charge $N=1$ ordinary monopole by taking \eqn{strng}{T_{1,2} = 0,
\, T_{3} = {\t} z.} To solve \rfn{drc} for  $\Psi$ we introduce
the operators $b, b^{\dagger}$: \eqn{cran}{ b =
{1\over{\sqrt{2\t}}} \left( {\p}_z +  x_3 + {\t} z
  \right), \quad \quad b^{\dagger} = {1\over{\sqrt{2\t}}} \left( -
{\p}_z +  x_3 + {\t} z  \right)\ , } which obey the oscillator
commutation relations: \eqn{osc}{ [ b, b^{\dagger} ] = [ c,
c^{\dagger} ] = 1\ .}Introduce  the {\it superpotential}
\eqn{sprn}{ W = x_3 z + {1\over 2}{\t} z^2\ , }so that  $b =
{1\over{\sqrt{2\t}}} e^{-W} {\p}_z e^{W}, \quad b^{\dagger} = -
{1\over{\sqrt{2\t}}} e^{W} {\p}_z e^{-W}$. Then equation \rfn{drc}
becomes: \eqn{drcnc}{\matrix{& b^{\dagger}  \Psi_{+} + c
{\Psi}_{-} = 0 \cr & c^{\dagger} {\Psi}_{+} -  b {\Psi}_{-} = 0
\cr} .}It is convenient to solve first the equation
\eqn{drcnch}{\matrix{& b^{\dagger} {\e}_{+} + c {\e}_{-} = 0\ ,
\cr & c^{\dagger} {\e}_{+} - b {\e}_{-} = 0 \ ,\cr}}with
${\e}_{\pm} (z, x_3) \in {\CH}$. The number of solutions to
\rfn{drcnch} depends on what the interval $I$ is. If $I = (a_{-},
a_{+})$ is finite then \rfn{drcnch} has the following solutions:
\eqn{slh}{\matrix{& {\ve}^{\a} \, = \, \pmatrix{{\e}_{+}^{\a} \cr
{\e}_{-}^{\a} \cr}\ , \, \a = 0, 1 \cr & {\ve}_{0}^{0} \, = \,
\pmatrix{ 0 \cr {1\over{\sqrt{{\z}_0}}} e^{-W} \vert 0 \rangle
\cr} \ ,  \qquad {\z}_0 = \int_{a_{-}}^{a_{+}} \, dz \, e^{-2W},
\, {\ve}_0^{1} = 0 \cr & {\ve}_{n}^{\a} \quad =  \quad \pmatrix{ b
\, {\b}_{n}^{\a} \vert n-1 \rangle \cr \sqrt{n} \, {\b}_{n}^{\a}
\vert n \rangle \cr},\,
 \qquad n > 0 \ ,\cr}}where $e_{0}, e_{1}$ will be the basis vectors in
 the two dimensional Chan-Paton space. The functions
 ${\b}_n^{0}, {\b}_{n}^{1}$
solve \eqn{schroed}{ \left( b^{\dagger}\, b + n \right)
{\b}_n^{\a} = 0,}and are required to obey the following boundary
conditions: \eqn{bndry}{\matrix{b{\b}_n^{1} (a_{+}) = 0, \quad &
\quad {\b}_{n}^{0} (a_{-}) = 0 \cr {\b}_{n}^{1} b {\b}_{n}^{1}
(a_{-}) = -1, \quad & \quad {\b}_{n}^0 b {\b}_{n}^0 (a_{+}) = 1
\cr}}which together with \rfn{schroed} imply that:\eqn{nrmeps}{
\int_{a_{-}}^{a_{+}}\, dz\, \left( {\ve}_{n}^{\a}
\right)^{\dagger} {\ve}_{m}^{\g} = {\d}^{\a\g} {\d}_{mn}, } A
solution to \rfn{drcnc} is given by: \eqn{asln}{\Psi = \sum_{n
\geq 0, \, {\a} = 0,1 } {\ve}_{n}^{\a} \cdot \langle n- {\a}
\vert\otimes e_{\a}^{\dagger}  .}and by virtue of \rfn{nrmeps} it
obeys \rfn{nrmcm}. All other solutions to \rfn{drcnc}, which are
normalizable on $I = (a_{-}, a_{+})$ are gauge equivalent to
\rfn{asln}.

If $I = (-\infty, 0)$ then the number of solutions to \rfn{drcnch}
is roughly halved. \eqn{slhuone}{\matrix{& {\ve} \, = \,
\pmatrix{{\e}_{+} \cr {\e}_{-} \cr}\ ,  \cr & {\ve}_{0} \, = \,
\pmatrix{ 0 \cr {1\over{\sqrt{{\z}_0}}} e^{-W} \vert 0 \rangle
\cr} \ ,  \qquad {\z}_0 = \int_{-\infty}^{0} \, dz \, e^{-2W}, \,
\cr & {\ve}_{n} \quad =  \quad \pmatrix{ b \, {\b}_{n} \vert n-1
\rangle \cr \sqrt{n} \, {\b}_{n} \vert n \rangle \cr},\,
 \qquad n > 0 \ ,\cr}} The functions ${\b}_n$
solve \eqn{schroeduone}{ \left( b^{\dagger}\, b + n \right) {\b}_n
= 0,}and are required to obey the following boundary conditions:
\eqn{bndryuone}{{\b}_{n} b {\b}_{n} (0) = 1, \qquad {\b}_n(z) \to
0, \, z \to - \infty}which together with \rfn{schroeduone} imply
that:\eqn{nrmepsuone}{ \int_{-\infty}^{0}\, dz\, \left( {\ve}_{n}
\right)^{\dagger} {\ve}_{m} = {\d}_{mn}, } A solution to
\rfn{drcnc} is given by: \eqn{aslnuone}{\Psi = \sum_{n \geq 0,}
{\ve}_{n} \cdot \langle n \vert .}and by virtue of
\rfn{nrmepsuone} it obeys \rfn{nrmcm}. All other solutions to
\rfn{drcnc} which are normalizable on $I = (-\infty, 0)$ are gauge
equivalent to \rfn{asln}.

\subsubsection{Generating solutions of the auxiliary problem: $U(1)$ case}
To get the solutions to \rfn{schroeduone} solve first the equation
for $n=1$ and then act on it by $b^{n-1}$ to generate the solution
for higher $n$'s. The result is: \eqn{betasuone}{{\b}_{n}(z) =
{{{\z}_{n-1}(x_3 + {\half}z)}\over \sqrt{{\z}_{n}(x_{3})
{\z}_{n-1}(x_{3})}}}where (we set $2\t = 1$):\eqn{zetas}{{\z}_n
(z) = \int_{0}^{\infty} \, p^n e^{2p z - {{p^2}\over 2}} \, {\rm
d}p}The functions ${\z}_n$ obey\ctn{grossnek}:
\eqn{zetauone}{\matrix{& {\z}_{n+1}(z) = 2z {\z}_n (z) +
n{\z}_{n-1}(z)\, , \cr & {\p}_z {\z}_n = 2{\z}_{n+1}\, ,  \cr &
{\z}_{n}(0) = 2^{{n-1}\over 2} \left( {n-1 \over 2} \right)!\cr}}
We have explicitly constructed ${\b}_n$ and thus $\Psi_\pm$, from
which we can determine, using  \rfn{sltn}, the Higgs and gauge
fields. To this end:

{\exc Show that: \eqn{ovrlp}{\int^{0}_{-\infty} {\b}_n b {\b}_n  =
{\z}_{n-1}{\z}_{n+1} - {\z}_{n}^2 = n{\z}_{n-1}^2 - (n-1)
{\z}_{n}{\z}_{n-2}\ , n > 0 }}Introduce the functions ${\xi},
{\tilde \xi}\ $ and ${\eta} = {\tilde\xi}^2$ :
\eqn{etaxi}{{\tilde\xi} ( n ) = \sqrt{{\z}_{n} \over {\z}_{n+1}},
\qquad {\eta} (n) = {{\z}_{n} \over{{\z}_{n+1}}}, \qquad {\xi} (n)
= \sqrt{n{\z}_{n-1} \over {\z}_{n}} \ . }We will need the
asymptotics of these functions for large $x_3$. Let $r^2_{n} =
x_3^2 + n$. For  $r_{n} + x_3 \to \infty$  we can estimate the
integral in \rfn{zetas} by the saddle point method. The saddle
point and the approximate values of ${\z}_n$ and $\eta_n$  are:
\bea & & {\bar p} \quad  =  \quad  x_3 + r_{n}\crt &   & {\z}_{n}
\quad   \sim  \quad \sqrt{{\pi}\over{r_{n}}}
 \left( x_3 + r_{n} \right)^{n + {1\over 2}} e^{{1\over{2}}
\left( x_3 + r_{n} \right)\left( 3x_3 - r_{n} \right)}\crt &  &
{\eta}_{n} \sim  \quad {1\over{x_3 + r_{n+1}}}  \left( 1 + {1\over
{4r^2_{n}}} + \ldots \right)  . \crt \label{sdlpnt} \eea\\ We
shall also need: \eqn{mstr}{{\z}_{0} (z) \sim \matrix{&
\sqrt{2\pi}e^{2z^2}, \quad z \to +\infty\cr &{\half}\vert z
\vert^{-1}, \quad z \to -\infty\cr}}

\subsubsection{Generating solutions of the auxiliary problem: $U(2)$ case}
It is easy to generate the solutions to \rfn{schroed}: first of
all, $$ f_{n}(z) = b^{n-1}(e^{W}) = e^{W(z)} h_{n-1}(2x_3 +z),
\qquad h_{k} (u) = e^{-{u^2 \over 4}} {d^{k}\over{{d u^k}}} e^{u^2
\over 4}$$is a solution . Then $$ {\hat f}_n = f_n (z)\int^{z}
{{du}\over{f_{n}(u)^2}}$$ is the second solution. Notice that, for
$k$ even, $h_{k} (u) > 0 $ for all $u$   and, for $k$ odd, the
only zero of $h_{k}(u)$ is at $u=0$, and $h_{k}(u)/u
> 0$ for all $u$. Therefore, ${\hat h}_k (z)$ is well-defined for
all $z$.

Consequently, \eqn{betas}{\matrix{& {\b}_{n}^{0}(z) = {\tilde\n}_n
\, f_n (z) \int_{a_{-}}^{z} {{du}\over{f_{n}(u)^2}} \ ,\cr &
{\b}_{n}^{1} = {\n}_{n}\,  \left( - {1\over{n f_{n+1}(z)}} + f_{n}
(z) \int^{a_{+}}_{z} {{du}\over{f_{n+1}(u)^2}} \right)\ , \cr} }
where \eqn{nus}{\matrix{{\n}_{n}^{-2} \quad = & \quad \left(
f_{n}(a_{-})f_{n+1}(a_{-}) \int_{a_{-}}^{a_{+}}
{{du}\over{f_{n+1}^2(u)}} - {1\over{n}} \right)
\int_{a_{-}}^{a_{+}} {{du}\over{f_{n+1}^{2}(u)}}\ , \cr
{\tilde\n}_{n}^{-2} \quad = &\quad \left(
f_{n}(a_{+})f_{n+1}(a_{+}) \int_{a_{-}}^{a_{+}}
{{du}\over{f_{n}^2(u)}} + 1 \right) \int_{a_{-}}^{a_{+}}
{{du}\over{f_{n}^{2}(u)}} \ .\cr }} (again, note that ${\b}^{\a}_n
(z)$ are regular at $z = - 2x_3$).

\subsection{Explicit $U(1)$ solution}
Now we present explicit formulae for the $U(1)$ monopole solution
and study its properties.
\subsubsection{Higgs/gauge fields}
The Higgs field is given by: \eqn{hgs}{{\Phi} =
\sum_{n=0}^{\infty} {\Phi}_{n} (x_3) \vert n \rangle \langle n
\vert \ , }it has   axial symmetry, that is commutes with the
number operator $c^{\dagger}c$. Explicitly: \bea {\Phi}_{n} \quad
& = & \quad {{{\z}_{n}}\over{{\z}_{n-1}}} -
{{{\z}_{n+1}}\over{{\z}_{n}}} = \, {\p}_{3} {\rm log} {\xi}_{n}
\crt  \, & = & (n-1) {\eta}_{n-2} - n {\eta}_{n-1}, \qquad n > 0
\crt \, & = & - {{\z}_1 \over {\z}_0} = -2x_3-{1 \over {\z}_0} ,
\qquad\quad n =0\, . \crt \label{hggs} \eea\\ {}To arrive at the
third line we used the fact that $$ {1\over \eta_n}-{1\over
\eta_{n+1}} = n \eta_{n-1} -(n+1) \eta_{n} \ , $$ which follows
immediately from the recursion relation for the $\z'$s in
\rfn{zetauone}. {}These fields are finite at $x_3=0$. Indeed as
$x_3 \to 0$, \eqn{hgszr}{ {\Phi}_{n} (x_3 = 0)\quad  = \sqrt{2}
\left( {{\left( {{n-1}\over 2} \right)!}\over{\left({{n-2}\over 2}
\right)!}} - {{\left( {{n}\over 2}
\right)!}\over{\left({{n-1}\over 2} \right)!}} \right) \ .} At the
origin: \eqn{orign}{{\Phi}_{0} (x_3 = 0)\quad = \quad  -
\sqrt{2\over \pi}\  .}
\ndt{}In the gauge where ${\Phi}$ is diagonal the component $A_3$
vanishes. In the same gauge the components $A_1, A_2$ (which we
consider to be anti-hermitian) are given by: \bea  & A_{c} =
{1\over 2} \left( A_1 + i A_2 \right), \quad  A_{c^{\dagger}} =
{1\over 2} \left( A_1 - i A_2 \right) = - A_{c}^{\dagger}
\nonumber \\ & A_{c} \quad = \quad {\xi}^{-1} [ {\xi}, c^{\dagger}
] = c^{\dagger} \left( 1 - {{\xi}(n) \over {\xi}(n+1)} \right)
\label{ggef} \eea\\ {}Again we see that the matrix elements of
$A_c$ are all finite and non singular.

\ndt{} From \rfn{ggef} we deduce: $$ F_{12} = 2\,
 \left( {\p}_{c}A_{c^{\dagger}} - {\p}_{c^{\dagger}} A_{c} + [ A_{c},
A_{c^{\dagger}} ] \right) = $$ \bea
 & 2 \, \left( \left[ {{{\xi}
(n)}\over{{\xi}(n+1)}} c, c^{\dagger} {{{\xi}
(n)}\over{{\xi}(n+1)}}  \right] - 1 \right) = \nonumber \\ & = 2
\, \sum_{n
> 0} \left( - 1 + (n+1) \left( {{{\xi}(n)}\over{{\xi}(n+1)}}
\right)^2 - n \left( {{{\xi} (n-1)}\over{{\xi}(n)}} \right)^2
\right) \vert n \rangle \langle n \vert + \nonumber \\ &
\qquad\qquad\qquad +   2\,  \left( - 1 + \left(
{{{\xi}(0)}\over{{\xi}(1)}} \right)^2 \right) \vert 0 \rangle
\langle 0 \vert  \  , \label{fonetwo} \eea from which it follows,
that: \bea  B_3 (n) \quad = \quad & 2 \left(1 - n
{{{\eta}_{n-1}}\over{{\eta}_{n}}}
   + \left(n-1\right)
{{{\eta}_{n-2}}\over{{\eta}_{n-1}}}
  \right)\nonumber \\
B_{c} \quad = \quad & {1\over 2} \left( B_1 + i B_2 \right) =
c^{\dagger} {{{\xi} (n) }\over{{\xi} (n+1) }} \left( {\Phi} (n) -
{\Phi} (n+1) \right)\ .\label{flds} \eea with the understanding
that at $n=0$: \eqn{batz}{  B_3(0) = 2\left(1 - {\z_1\over
\z_0^2}\right) .   }

\subsubsection{Instantons, monopoles, and Yang ansatz}
\ndt As in the ordinary gauge theory case the monopoles  are the
solutions of  the instanton equations in four dimensions, that are
invariant under translations in the fourth direction $x_4$.  We
observe that the solution  presented above \rfn{hggs},\rfn{ggef},
can also be cast in the Yang form: Take $\xi = \xi (x_3, n)$ as in
\rfn{ggef}. Then ${\p}_3 {\xi}$ commutes with ${\xi}$ and we can
write ${\p}_3 {\xi} {\xi}^{-1} = {\p}_{3} {\rm log} {\xi}$. The
formulae \rfn{ynga} yield exactly \rfn{ggef} and \rfn{hggs} with
${\Phi} = i A_{4}$.

\subsubsection{All of the above and Toda lattice}

\ndt At this point it is worth mentioning the relation of the
noncommutative Bogomolny equations with the Polyakov's non-abelian
Toda system on the semi-infinite one-dimensional lattice. Let us
try to solve the equations \rfn{bgmln} using the Yang ansatz and
imposing the axial symmetry: we assume that ${\xi} ( x_1, x_2, x_3
) = {\xi} (n , x_3)$, $n = c^{\dagger}c$. Then the equation
\rfn{ynge} for the $x_4$-independent fields reduces to the system:
\eqn{toda}{ {\p}_{t} ( {\p}_{t} g_n g_n\inv )  -
g_{n}g_{n+\scriptscriptstyle 1}\inv + g_{n-\scriptscriptstyle 1}
g_{n}\inv = 0} where $$ g_{n}(t) = {{e^{{t^2 \over 2}}}\over{n!}}
{\xi}^2 \left(n, {t\over 2}\right)\ , $$ (notice that $g_{n}(t)$
are ordinary matrices). In the $U(1)$ case we can write $$
g_{n}(t) = e^{{\a}_{n}(t)}\ , $$ and rewrite \rfn{toda} in a more
familiar form: \eqn{todai}{{\p}_{t}^2 {\a}_{n} + e^{{\a}_{n-1} -
{\a}_{n}} - e^{{\a}_{n} - {\a}_{n+1}} = 0}For $n = 0$ these
equations also formally hold if we set $g_{-1} = 0$ (this boundary
condition follows both from the Bogomolny equations and the same
condition is imposed on the Toda variables on the lattice with the
end-points).

Our Higgs field ${\Phi}_{n}$ has a simple relation to the
${\a}$'s: $$ {\Phi}(x_3, n) = - 2x_3 + {\a}_{n}^{\prime} (2x_3) \
. $$ Our solution to \rfn{todai} is: \eqn{ours}{{\a}_{n} = {1\over
2} t^2 + {\rm log} \left( {{n{\z}_{n-1}(t/2)}\over{{\z}_{n}(t/2)}}
\right) - {\rm log} ( n! )\ .}

It is amusing that Polyakov's motivation for studying   the system
\rfn{toda} was  the structure of loop equations for   lattice
gauge theory. Here we encountered these equations in the study of
the {\it continuous}, but noncommutative, gauge theory, thus
giving more evidence for their similarity.

We should note in passing that in the integrable non-abelian Toda
system one usually has two `times' $t, {\bar t}$, so that the
equation \rfn{toda} has actually the form \ctn{mtoda}:
\eqn{todaii}{ {\p}_{t} ( {\p}_{\bar t} g_n g_n\inv )  -
g_{n}g_{n+\scriptscriptstyle 1}\inv +  g_{n-\scriptscriptstyle 1}
g_{n}\inv = 0\ .}

It is obvious that these equations describe four-dimensional axial
symmetric instantons on the noncommutative space with the
 coordinates $t, {\tb}, c, c^{\dagger}$ of which only half
is noncommuting.

\subsubsection{The  mass of  the  monopole}

\ndt{} In this section we restore our original units, so that
$2{\t}$ has dimensions of (length)$^2$. From the formulae
\rfn{sdlpnt} we can derive the following estimates: \eqn{asmphi}{
{\Phi} (n) \sim -{1\over{2r_{n}}}= -{1\over{2\sqrt{x_3^2+2 {\t}
n}}}\quad n \neq 0, \, r \to \infty \ .} Instead, for $n=0$ we
have: \bea & {\Phi} (0) \sim - {{x_3}\over{\t}}, \qquad x_3 \to +
\infty \nonumber \\ &{\Phi} (0) \sim -{1\over{2\vert x_3 \vert}},
\qquad x_3 \to - \infty \ .\label{asm} \eea {}The  asymptotics of
the magnetic field is clear from the Bogomolny equations and the
behaviour of $\Phi$. Thus, for example, \eqn{asyb}{B_3(n) =
-\partial_3 \Phi(n) = -{x_3\over 2r_{n}^3} ,\quad n\neq0  , } and
similarly for the other components of $B$. This is easily
translated into ordinary position space, since, for large $n$,    $B_i(n,
x_3) \sim B_i( x_1^2+x_2^2 \sim n, x_3)$. Therefore the magnetic
field for large values of $x_3 $ and $n$, or  equivalently large
$x_i$ is that of a pointlike magnetic charge at the origin.
However the $n=0$ component of $B_3$ behaves differently for large
positive $x_3$: \eqn{Bthras}{ B_3(n=0) = -\p_3\Phi(0) =
{1\over{\t}} \ .} Notice, that this is exactly the value of the
$B$-field. Thus, in addition to the magnetic charge at the origin
we have a flux tube,
 localized in a Gaussian packet in the $(x_1 ,x_2)$ plane, of the
size $\propto \theta$, along the positive $x_3$ axis. The monopole
solution is indeed a smeared version of the Dirac monopole,
wherein the Dirac string (the D-string!) is physical.

To calculate the energy of the monopole we use the  Bogomolny
equations to reduce the total energy to a boundary term: \bea &
{\CE} = {1\over{2g_{\rm YM}^2}} \, \int d^3 x \, \left( {\vec B}
\star {\vec B} + {\vec\nabla} {\Phi} \star {\vec\nabla} {\Phi}
\right)
  = \nonumber \\
& {1\over{2g_{\rm YM}^2}} \, \int d^3 x \left( {\vec B} +
{\vec\nabla} \Phi \right)^2  - {1\over{2g_{\rm YM}^2}} \, \int d^3
x \, {\vec\nabla} \cdot \left( {\vec B} \star {\Phi} + {\Phi}
\star {\vec B} \right) =  \nonumber \\ & \nonumber \\ &  {{2\pi
{\t}}\over{2g_{\rm YM}^2}} \int dx_3 \sum_{n} \langle n \vert
{\p}_3^2 {\Phi}^2 + 4 {\p}_{c} \left( {\xi}^2 \left(
{\p}_{c^{\dagger}} {\Phi}^2 \right) {\xi}^{-2}\right) \vert n
\rangle \ , \label{msm} \eea\\ where in the last line we switched
back to the Fock space. Here we meet the \nct boundary term,
discussed in the section $2$. Let us choose as the infrared
regulator box the ``region'' where $\vert x_3 \vert \leq L, 0 \leq
n \leq N$, $L \sim \sqrt{2{\t}N} \gg 1$. With the help of
\rfn{bndryt} the total integral in \rfn{msm} reduces to the sum of
two terms (up to the factor ${\pi {\t}}\over{g_{\rm YM}^{2}}$):
\bea & 4 N \int_{-L}^{L} dx_3 \, {{{\eta}_{N-1}}\over{{\eta}_{N}}}
\left( {\Phi}_{N}^2 - {\Phi}_{N+1}^2 \right)\cr & \qquad \qquad
+\sum_{n=0}^{N} {\p}_3 {\Phi}_{n}^2 \vert^{x_3 = +L}_{x_3 = - L} \
. \label{msmi} \eea The first line in \rfn{msmi} is easy to
evaluate and it vanishes in $L \to \infty$ limit. The second line
in \rfn{msmi} contains derivatives of the Higgs field evaluated at
$x_3 = L \gg 0$ and
 at $x_3 = -L \ll 0$. The former is estimated using the
$z \gg 0$ asymptotics in\rfn{sdlpnt}, and produces: $$
\sum_{n=0}^{N} {\p}_3 {\Phi}^{2}_{n} (x_3 = L) \sim
{{2{\t}(N-1)}\over{L^3}} + 2{{L}\over{{\t}^2}} $$ The diverging
with $L$ piece comes solely from the $n=0$ term. Finally, the $x_3
= -L$ case is treated via $z \ll 0$ asymptotics in \rfn{mstr}
yielding the estimate $\sim {\t}N/L^{3}$ vanishing in the limit of
large $L, N$.

Hence the total energy is given by \eqn{ttlms}{ {\CE} \propto
{{2\pi {\t} \times 2 L}\over{2g_{\rm YM}^2 {\t}^2}} = {2{\pi} L
\over{g_{\rm YM}^2 {\t}}} \ ,} which is the mass of a string of
length $L$ whose tension is \eqn{tnsni}{T = {{2{\pi}\over{g_{\rm
YM}^2 {\t}}}} \ .}

\subsubsection{Magnetic charge}

It is instructive to see what is the magnetic charge of our
solution. On the one hand, it is clearly zero: \eqn{chrn}{ Q
\propto \int_{{\p} ({\rm space})} {\vec B} \cdot d{\vec S} = \int
d^3 x \,\,  {\vec\nabla} \cdot {\vec B} = 0 } since the gauge
field is everywhere non-singular. On the other hand, we were
performing a $\t$-deformation of the Dirac monopole, which
definitely had magnetic charge. To see what has happened let us
look at \rfn{chrn} more carefully. We  again introduce the box and
evaluate the boundary integral \rfn{chrn} as in \rfn{trcrg}:
\eqn{chnri}{{Q\over{2\pi}} = \sum_{n=0}^{N} \left[ B_{3} (x_3 = L,
n) - B_{3} (x_3 = -L, n) \right] + 4N \int_{-L}^{L} dx_3
{{\eta}_{N-1} \over {\eta}_{N}} \left( {\Phi}_{N} - {\Phi}_{N+1}
\right)} It is easy to compute the sums \bea & \qquad \qquad
\qquad \sum_{n=0}^{N} B_{3} (x, n) =  {\p}_3 {{\z}_{N+1} \over
{\z}_{N}} \cr & 4 N \int_{-L}^{L} dx_3 {{\eta}_{N-1} \over
{\eta}_{N}} \left( {\Phi}_{N} - {\Phi}_{N+1} \right) =
 4 (N+1) \int {{\xi}_{N}^2 \over {\xi}_{N+1}^2} \, d \, {\rm log}
  {{\xi}_{N}
\over {\xi}_{N+1} } = \cr & = 2 (N+1) \left( {{\xi}_{N} \over
{\xi}_{N+1}} \right)^2     \vert_{x_3 = -L}^{x_3 = + L}  = \quad 2
N {{{\z}_{N-1} {\z}_{N+1}}\over{{\z}_{N}^2}} \vert_{x_3 = -L}^{x_3
= + L}\ ,  \label{sms} \eea {}and the total charge vanishes as:
\eqn{ttlch}{Q = \left[ 2N {{{\z}_{N-1}
{\z}_{N+1}}\over{{\z}_{N}^2}}
 +
{\p}_3 \left( {{\z}_{N+1} \over {\z}_{N}} \right)
 \right]_{x_3 = -L}^{x_3 = + L} \equiv 2(N+1)
 \vert_{x_3 = +L} - 2(N+1) \vert_{x_3 = -L} \ . }

{}We can better understand the distribution of the magnetic field
by looking separately at the fluxes through the ``lids'' $x_3 =
\pm L$ of our box and through the ``walls'' $n = N$.

{}The walls contribute $$ \left[ 2N {{{\z}_{N-1}
{\z}_{N+1}}\over{{\z}_{N}^2}} \right]_{x_3 = -L}^{x_3 = + L} \sim
- {{L}\over{\sqrt{L^2 + N}}} \sim - 1 \ , $$ while the lids
contribute $\sim + 1$. Let us isolate the term $B_{3}( +L, n=0)
\to +2$ (recall \rfn{asm}). It contributes to the flux through the
upper lid. The rest of the flux through the lids is therefore
$\sim -1$. Hence the flux through the rest of the ``sphere at
infinity'' is $-2$ and it is roughly uniformly distributed ($-1$
contribute the walls and $-1$ the lids).  So we get a picture of a
spherical magnetic field of a monopole together with a flux tube
pointing in one direction.

{}This spherical flux becomes observable in the naive ${\t} \to 0$
limit, in which the string becomes localized at the point $x_3 =
0$, $n=0$ (since the slope of the linearly growing ${\Phi}_{0}
\sim {x_3 \over {\t}}$ becomes infinite). In the $\t = 0$ limit we
throw out this point and all of the string.

Thus we found an  explicit analytic expression for a soliton in
the U(1) gauge theory on a noncommutative space. The solution
describes    a magnetic monopole  attached to a finite
  tension string, that runs off to infinity tranverse to the
noncommutative plane. This soliton has a clear reflection in type
IIB string theory. If the gauge theory is realized as an ${\ap}
\to 0$ limit of the theory on a D3-brane in the IIB string theory
in the presence of a background NS B-field,
 then the monopole with the string attached is nothing
but the D1-string ending on the D3-brane. What is unusual about
the solution that we found is that it describes this string as a
non-singular field configuration. Moreover, one can show that the
long wavelength fluctuations of this string are described by the
gauge theory \ctn{grossneki,grossnekii}.

\subsection{Noncommutative U(2) monopole.}

Now we shall describe the \nct version of the 't Hooft-Polyakov
monopole\ctn{thpol}.

We are interested in the $U(2)$ gauge theory on the \nct three
dimensional space. Recall that $H \approx {\bf C}^2$ is the
Chan-Paton space, i.e. the fundamental representation for the
commutative limit of the gauge group, and let $e_{0}, e_{1}$
denote an orthonormal basis in $H$. The \nct version of the
fundamental representation is infinite dimensional, isomorphic to
$H \otimes {\CH}$. That is, the $U(2)$ matter fields ${\Psi}$
belong to the space ${\CH} \otimes Fun (x^3) \otimes \left( {\CH}
\otimes H \right)$, where the first two factors make it a
representation of the algebra ${\CA}_{\t}$ of \nct functions on
${\IR}^3$, while the second two factors make it a representation
of the $U(2)$ \nct gauge group. Actually, the latter is isomorphic
to the group of ($x^3$-dependent) unitary operators in the Hilbert
space ${\CH} \otimes H$. Now, the Hilbert space ${\CH} \otimes H$
is isomorphic to ${\CH}$ itself: \eqn{iso}{ \vert n \rangle
\otimes \, e_{\a} \leftrightarrow \vert 2n + {\a} \rangle\ .} Now
the solutions can (and will) have a finite non-trivial magnetic
BPS charge: \eqn{mgntch}{Q_{m} = \int dx^3 {\Tr}_{\CH} {\p}_i
\left( {\Tr}_{H} {\Phi} B_i \right)\ , } where $$ B_i = {{\i}\over 2}
{\ve}_{ijk} [D_j, D_k] - {\d}_{i3} {\t} \ ,$$ and the Higgs field
${\Phi}$ approaches $$ \pmatrix{ a_+ & 0 \cr 0 & a_{-}} \otimes
{\rm I}_{\CH}$$ as $x_3^2 + 2{\t} c^{\dagger} c \to \infty$.

\ndt In \rfn{asln} we already found a two-component spinor
vector-function $${\Psi} (z, {\vec x}) = \pmatrix{ {\Psi}_{+} \cr
{\Psi}_{-} }\ , $$ which obeys the equation \rfn{drc}. The
solution to \rfn{drc} is defined up to   right multiplication by
an element of ${\rm Mat}_{2}({\CA}_{\t}) \approx {\CA}_{\t}
\otimes {\rm End}(H)$: ${\Psi} \mapsto {\Psi} u$. Among these
elements the unitary elements (i.e. the ones which solve the
equation $uu^{\dagger} = u^{\dagger}u = 1$) are considered to be
the gauge transformations. In the commutative setup one normalizes
${\Psi}$ as follows: \eqn{nrmcm}{\int \, dz \, {\Psi}^{\dagger}
{\Psi} = {\rm I}_{2} = \pmatrix{ 1 & 0 \cr 0 & 1}\ .}The
normalization \rfn{nrmcm} only implies that the transformations
$u$ must obey: $u^{\dagger}u = 1$. The finite matrices $u$ would
then automatically obey $uu^{\dagger} =1$. However, in the
infinite-dimensional case this is not true. The operator
$uu^{\dagger}$ is merely a projection, which may have a kernel.
The discussion on whether such projections should be viewed as
gauge transformations in \nct gauge theory can be found in
\ctn{fkii,fkiii}.
\subsubsection{Higgs/gauge field}
Finally, given ${\Psi}$ the solution for the gauge and Higgs
fields is given explicitly by \rfn{sltn} where now one integrates
over $z$ from $a_{-}$ to $a_{+}$. We now are in position to
calculate the components of the Higgs field and of the gauge
field. We start with \eqn{hggsi}{\matrix{& {\Phi} = \int \, dz \,
z \, {\Psi}^{\dagger} {\Psi} = \cr & \quad \sum_{n \geq 0, \,
{\a},{\g} = 0,1} {\vf}^{\a\g}_{n} \cdot e_{\a} e_{\g}^{\dagger}
\otimes \vert n - {\a} \rangle \langle n - {\g} \vert \, , \cr &
{\rm where} \quad {\vf}^{\a\g}_{n} = \int \, dz \, z
{\ve}_{n}^{\a, \dagger} {\ve}_{n}^{\g} = - 2x_3 {\d}^{\a\g} + \int
(b{\b}_n^{\a}) (b + b^{\dagger}) (b{\b}_n^{\g}) + n {\b}_n^{\a} (b
+ b^{\dagger}) {\b}_n^{\g}  \cr  & \qquad\qquad = -
2x_3{\d}^{\a\g}  + \left( (b{\b}_n^{\a})(b{\b}_n^{\g}) - n
{\b}_n^{\a}{\b}_n^{\b}\right)\vert_{a_{-}}^{a_{+}} \ .\cr}} The
component $A_3$ of the gauge field vanishes, just as in  the case
of the U(1) solution of \ctn{grossnek}: \eqn{athree}{A_3 = \int
{\Psi}^{\dagger} {\p}_3 {\Psi} = \int \left( (b {\b}_n^{\a})
{\p}_3 ( b {\b}_n^{\g} ) + n {\b}^{\a}_n {\p}_3 {\b}_n^{\g}
\right) \cdot e_{\a} e_{\g}^{\dagger} \otimes \vert n - {\a}
\rangle \langle n - {\g} \vert = {\half} {\p}_3 \int
{\Psi}^{\dagger} {\Psi} = 0 \ .}

The components $A_1, A_2$ can be read off the expression for the
operator $D$: \eqn{dopr}{\matrix{& D = -\int  \, dz \,
{\Psi}^{\dagger} c^{\dagger} {\Psi} \cr  & = \sum_{n \geq 0,
{\a},{\g} = 0,1} D_{n}^{\a\g} \cdot \, e_{\a} e_{\g}^{\dagger}
\otimes \vert n+1 - {\a} \rangle \langle n - {\g} \vert\ , \cr &
{\rm where} \, \, D_{n}^{\a\g} =  - \sqrt{n} \left(
{\b}_{n+1}^{\a} (b{\b}_{n}^{\g}) \right) \vert_{a_{-}}^{a_{+}}\ .
\cr}}

The solution \rfn{hggsi},\rfn{dopr} has several interesting length
scales involved (recall that our units above are such that $2{\t}
=1$): $$ {\t} \vert a_{+} - a_{-} \vert, \quad \sqrt{\t}, \quad
{1\over{\vert a_{+} - a_{-} \vert}\ .} $$

By shifting $x_3$ we can always assume that $a_{-} = 0, a_{+} = a
> 0$.

\subsubsection{Suspended D-string}

In this section we set $\t$  back to $\half$. As discussed
in\ctn{grossnekii} the spectrum of the operators $D_A$, $A =
1,2,3,4$ determines the ``shape'' of the collection of D-branes
the solution of the generalized IKKT model\ctn{cds} corresponds
to. To ``see'' the spatial structure of our solution let us
concentrate on the $\langle 0 \vert {\Phi} \vert 0 \rangle$ piece
of the Higgs field, for it describes the profile of the D-branes
at the core of the soliton. From \rfn{hggsi} we see that $$\langle
0 \vert {\Phi} \vert 0 \rangle = \pmatrix{{\rho}_{+} & 0 \cr 0 &
{\rho}_{-} \cr} \ , $$ where ${\rho}_{+} = {\vf}^{00}_{0},
{\rho}_{-} = {\vf}^{11}_{1}$.

Let us look specifically at the component ${\rho}_{+}$ of the
Higgs field: \eqn{snakeyes}{\matrix{{\rho}_{+} \, &= \, - {\half}
{{\p}\over{{\p} x_3}} {\rm log} \left( \int_{0}^{a} dp \, e^{-2x_3
p - {\half} p^2} \right)\ ,\cr & = - 2x_3 + \langle\langle p
\rangle\rangle_{2x_3}^{a + 2x_3}\ , \cr & = \, - 2x_3 - 2
{{e^{-{{(a + 2x_3)^2}\over 4}} - e^{-{{(2x_3)^2}\over
4}}}\over{{\g}( {\scriptstyle a + 2x_3}) - {\g}( {\scriptstyle
2x_3})}} \ ,\cr }} where \eqn{defs}{ \langle\langle {\CO}
\rangle\rangle_{\a}^{\b} = {{\int_{\a}^{\b} {\CO} e^{-{p^2 \over
4}} \, dp}\over{\int_{\a}^{\b} e^{-{p^2 \over 4}} \, dp}} ,  \quad
{\g}(z) = \int_{0}^{z\over 2} dp \, e^{-{p^2 \over 4}}\ .  }
   The
$\langle\langle\ldots\rangle\rangle$ representation of the answer
helps to analyze the qualitative behavior of the profile of
$\rho_{+}$. Clearly, the truncated Gaussian distribution which
enters the expectation values $\langle\langle\ldots\rangle\rangle$
in \rfn{snakeyes} favors $p \approx 0$ if ${\a} < 0 < {\b}$, $p
\approx {\a}$ for ${\a} > 0$ and $p \approx {\b}$ for ${\b} < 0$.
Thus, \eqn{spke}{\matrix{{\rho}_{+} \sim 0, & \qquad x_3 > 0 \cr
{\rho}_{+}\sim - 2x_3, & \qquad
   0 > x_3 > {-\half} a \cr
  {\rho}_{+} \sim a, & \qquad {-\half} a > x_3 \ . \cr}}
This behavior agrees with the expectations about the tilted
D1-string suspended between two D3-branes separated by a distance
$\vert a \vert$. The eigenvalue ${\rho}_{+}$ corresponds roughly
to the the transverse coordinate of the D1 string, that runs from
$a$ at large negative $x_3$ to $0$ at large positive $x_3$. In
between the linear behavior of the Higgs field corresponds to the
D1 string tilted at the critical angle. Indeed, for large $a \gg
1$,  in the region $0
> x_3
> {-\half} a$ this solution looks very similar to that of a single
fluxon \ctn{grossneki}. Another eigenvalue of $\langle 0 \vert
{\Phi} \vert 0 \rangle$, ${\rho}_{-}$, is given by:
\eqn{romin}{\matrix{& {\rho}_{-} = {{2x_3 (2x_3 + a) M + M^2 -
(2x_3 + a)^2 - e^{-2x_3 a - {{a^2}\over{2}}}}\over{M ( 2x_3 + a -
2x_3 M)}} \cr {\rm where} &  M = e^{-2x_3 a - {{a^2}\over{2}}} +
(2x_3 + a) \int_{0}^{a} e^{-2x_3 p - {{p^2}\over{2}}} \, dp \cr}}
At this point, however, we should warn the reader that only the
eigenvalues of the full, $2 \infty \times 2 \infty$ operator
${\Phi}$ should be identified with the D-brane profile. The
components ${\rho}_{\pm}$ do not actually coincide with any of
them. The eigenvalues of ${\Phi}$, as it follows from the
representation \rfn{hggs}, are located between $0$ and $a$, which
is also what we expect from the dual D-brane picture
\ctn{hashimoto}.

\subsection{Tension of the monopole string versus that of D-string}

In this section we shall match the tension of the string we
observed in the $U(1)$ monopole solution to that of D-string
ending on the D3-brane in the presence of the constant $B$-field.
As already mentioned, a D-string ending on a D3-brane in the
presence of the constant $B$-field bends. To analyze this bending
one could use the exact solution of the Dirac-Born-Infeld theory
\ctn{moriyama}, the $B$-deformed spike solutions of
\ctn{curtjuan}. However, for our qualitative analysis, it is
sufficient to look at the linearized equations. If we replace the
DBI Lagrangian by its Maxwell approximation, then the BPS
equations in the presence of the $B$-field will have the form:
\eqn{bpsbf}{B_{ij} + F_{ij} + \sqrt{{\rm det}g} \, {\ve}_{ijk} \,
g^{kl}{\p}_{l}{\Phi} = 0 \ , } where we should use the closed
string metric \rfn{clstr}. The solution of \rfn{bpsbf} is:
\eqn{slntn}{{\Phi} = B \left( 1 + \left(
{{\t}\over{2{\pi}{\a}^{\prime}}} \right)^2 \right) x_3 -
{1\over{2r}}, \qquad r^2 = x_3^2 + {1\over{\left( 1 + \left(
{{\t}\over{2{\pi}{\a}^{\prime}}} \right)^2 \right)}} \left( x_1^2
+ x_2^2 \right) \ .} The linearly growing piece in ${\Phi}$ should
be interpreted as a global rotation of the D3-brane, by an angle
${\psi}$, ${\rm tan}{\psi} = {{\t}\over{(2{\pi}{\a}^{\prime})}}$.
This conclusion remains correct even after the full non-linear BPS
equation is solved (see \ctn{moriyama}.  Notice however that we
fix $G_{ij} = {\d}_{ij}$ instead of $g_{ij} = {\d}_{ij}$ as in
\ctn{moriyama}). The singular part of ${\Phi}$, the spike,
represents the D-string. If we rotate the brane, then the spike
forms an angle ${{\pi}\over{2}} -{\psi}$ with the brane.  If we
project this spike on the brane, then the energy, carried by its
shadow per unit length, is related to the tension of the D-string
via: \eqn{eft}{{{T_{D1}}\over{{\rm sin}{\psi}}} =
{1\over{2{\pi}{\a}^{\prime} g_{s}}} {{\sqrt{ (
2{\pi}{\a}^{\prime})^2 + {\t}^2}}\over{{\t}}} = {{(
2{\pi}{\a}^{\prime})^2 + {\t}^2}\over{2{\pi} g_{\rm YM}^2
({\a}^{\prime})^2 {\t}}} \, .}However, this is not the full story.
The endpoint of the D-string is a magnetic charge, which
experiences a constant force, induced by the background magnetic
field. If we had introduced a box $-L \leq x_3 \leq x_3$ of the
extent $2L$ in the $x_3$-direction, then in order to bring a
tilted D-string into our system from outside $x_3 > L$ of the box
to $x_3 = 0$ we would have had to spend an energy equal to
${T_{D1} \over{{\rm sin}{\psi}}} L$, but we would have been helped
by the magnetic force, which would decrease the work done
by\ctn{hashimoto} $$ {{2\pi L}\over{g_{\rm YM}^2}} B_3 = {{2\pi
L}\over{g_{\rm YM}^2}} B_{12} \,\, g^{11} g^{22} \sqrt{g} \sim
{1\over{({2\pi} {\a}^{\prime} g_{\rm YM})^2}} {\t} \, . $$In sum,
the energy of the semi-infinite D-string in the box  per unit
length in the $x_3$ direction will be given by
\eqn{efti}{{{2{\pi}}\over{g_{\rm YM}^2 {\t}}} \, .} This
expression coincides with our tension \rfn{tnsni}. On dimensional
grounds, non-commutative gauge theory cannot produce any other
dependence of the tension on ${\t}$ but that given  in
\rfn{tnsni}.

\section{Conclusions and historic remarks}

To conclude these lectures I would start with a very short survey
of the topics not included into them. The abovementioned
factorization of the OPE algebra of the open string vertex
operators was argued to be useful in analyzing various instability
issues, condensation of tachyons, responsible for the decays of
the unstable D-branes\ctn{tachyonsen,tachyons}. Moreover, \nct
gauge theories allow to see both (many of) the non-BPS D-branes as
classical solutions and the processes of their decays
\ctn{amgs,klh,grossnekii}.

My interest in \nct theories was prompted by the work with
V.~Fock, A.~Rosly and K.~Selivanov in 1991 on the geometric
quantization of higher-dimensional dimensional Chern-Simons
theories\ctn{wwthink}, where we encountered moduli spaces of
codimension two four dimensional foliations with flat $U(1)$
connections on the fibers as the classical phase space of the five
dimensional theory (in modern terms these would be supersymmetric
cycles, if they were holomorphic). Previously thee theories were
encountered in the context of gauge anomalies in\ctn{fadsha}. An
example of such foliation would be an irrational foliation of a
four-torus, the space suited for the analysis by \nct geometry. In
1994 with G.Moore and S.~Shatashvili I tried to understand
S-duality of the ${\CN}=4$ gauge theory on ALE manifolds
following\ctn{vafawitten} and then together with A.~Losev we came
from that to the attempt of constructing the four dimensional
analogue of the two dimensional RCFT\ctn{avatars}. In the course
of this study we realized that the construction
of\ctn{nakajimaale} of the instanton moduli spaces on ALE
manifolds, whose Euler characteristics used in\ctn{vafawitten}
were generated by modular forms actually gave nontrivial answers
already in the $U(1)$ case, which is almost impossible to achieve by 
the
ordinary gauge fields. With the help of I.~Grojnowski we realized
that Nakajima constructed not the instantons but the torsion free
sheaves, who had no gauge theory interpretation (but were used by
algebraic geometers to construct compactifications of instanton
moduli spaces, e.g. Gieseker compactification). For some time
these moduli spaces were a mystery. This mystery was getting
deeper after discovery of D-branes by J.~Polchinski\ctn{polch} and
the realization by E.~Witten and M.~Douglas\ctn{brinst} that the
D(p-4)-branes within Dp-brane are instantons. Now, the D4-brane
carries only a $U(1)$ gauge field on it, so what is the D0-brane
which is dissolved inside? This question is hard to ask if the
gauge theory is not decoupled from the rest of the ten dimensional
string theory, but it was soon realized that a constant $B$-field
may help (in the similar M-theory context turning on $C$-field
helped to decouple fivebrane from the eleven dimensional sugra
modes\ctn{abkss}). Things came together after we had a very
fruitful lunch with A.~Schwarz in ITP, Santa Barbara in the spring
of 1998, where he explained to me his paper with A.~Connes and
M.~Douglas, and I explained to him what we have learned about
torsion free sheaves with other authors of\ctn{avatars}. After
brief discussion we became confident that the gauge fields on the
\nct ${\IR}^4$ must be {\it i)} constructed with the help of ADHM
construction applied to the deformed ADHM
equations\ctn{nakajimares}, which was shown by Nakajima to
parameterize torsion free sheaves\ctn{nakajima} on ${\IC}^2$; {\it
ii)} obey instanton equations on the \nct space. It was a matter
of simple algebraic manipulations to check that this was the
case\ctn{neksch}. But then the devil of doubts started to hunt me.
Several people asked us whether the $U(1)$ instantons were
non-singular. The relation to large $N$ gauge theory suggested
they didn't exist, for the instanton effects usually die out in
the large $N$ limit. Explicit computations\ctn{branek,fki} seemed
to imply that the $U(1)$ instantons had to live over the space of
complicated topology. Also, with D.~Gross we were finding
solutions to the BPS equations in three dimensions which seemed to
follow from ADHM ansatz and yet did not solve Bogomolny equation
everywhere in the space. Later on we realized that these solutions
were obtained from the true solution by applying the
``generalized'' gauge transformation $u$, discussed below the
equation\rfn{nrmcm}, and which actually could produce a source.
But in the spring of 2000 I was not yet aware of that and during
the talk at CIT-USC center I was forced by E.~Witten to announce
that \nct instantons existed not on \nct ${\IR}^4$ but on some
space, obtained from ${\IR}^4$ by a sequence of blowups, whose
commutative limit was described in \ctn{branek}. This point of
view was immediately criticized by E.~Witten himself, for it was
not consistent with many results on the counting of states of
D-branes. Then a week later D.~Gross and I found true sourceless
solution to the monopole equations\ctn{grossnek}.  And later on I
realized that the explicit formulae for the $U(1)$ instanton gauge
field presented in \ctn{neksch} were harmed by the same plague:
they were written in the singular gauge. Almost at the same time
this conclusion was reached by K.~Furuuchi\ctn{fkiii,nekinst}.

Independently of all this story, it is natural to look for the
mechanism of the confinement in the \nct gauge theory, hoping to
transport these results to the large $N$ ordinary gauge theory. We
find\ctn{grossnekii} all sorts of magnetically charged objects in
the gauge theory, whose mass/tension goes to zero in the limit
${\t} \to \infty$. Whether this could provide the sought for
mechanism for the confinement via condensation of magnetic charges
still remains to be seen.

As far as other extensions of the work reported above are
concerned let us mention a few. First of all, it is quite
interesting to construct \nct instantons on other spaces, for
example ALE manifolds\ctn{neksch,calin}, tori or K3
manifolds\ctn{torik}, or Del Pezzo surfaces or cotangent bundles
to curves (see \ctn{nakajimaii}). In the latter three cases the
very notion of the underlying \nct manifold is missing (for
orbifold K3's one can make orbifolds of \nct tori,
though\ctn{nctorb}). One could also try to construct
supersymmetric gauge field configurations in higher dimensions
(see \ctn{nakajimaiii} for some relevant algebraic geometric
results). Also, $SO(n)$ and $Sp(n)$ theories \ctn{shakhino} are
quite curious to look at. The construction of the ordinary
instantons\ctn{brinst} in these theories with D-brane techniques
shows many interesting surprises\ctn{quivers}. It is also quite interesting
to generalize the \nct twistor approach of \ctn{twistor} to more 
general spaces.
Another subject is
the usage of the gauge theory/string duality in the form of
supergravity duals of \nct gauge theories\ctn{ncsugra}. One can
construct many sugra duals of the solitons in \nct theories and
study their strong coupling behaviour\ctn{shakhin}. Yet another
interesting topic is the construction of electrically charged
solitons. They are important in the realizing of closed strings
within open string field theories in the lines of
\ctn{shatashvili}.

\section*{Acknowledgements}

I thank all the people mentioned above and also H.~Braden, S.~Cherkis,
P.-M.~Ho, I.~Klebanov, S.~Majid, J.~Maldacena, A.~Polyakov,
D.~Polyakov, M.~M.~Sheikh-Jabbari, for various stimulating
discussions and questions.

I am especially thankful to my wife Tatiana Piatina for her
understanding and support.

The research was partially supported by NSF under the grant
PHY94-07194, by Robert H.~Dicke fellowship from Princeton
University, partly by RFFI under grant 00-02-16530, partly by the
grant 00-15-96557 for scientific schools. I thank the organizers
of the Trieste spring school on D-branes and strings for the kind
invitation to present the results reported here. I thank the ITP
at University of California, Santa Barbara, CIT-USC center for
Theoretical Physics, String Theory Group at Duke University,
Universities of Barcelona and Madrid, and Isaac Newton Institute
for Mathematical Sciences (Cambridge, UK) for hospitality while
these lectures were written up.

\end{document}